\documentclass[aps,pre,preprint,groupedaddress]{revtex4-1}
\usepackage[T1]{fontenc}
\usepackage[latin9]{inputenc}
\usepackage{amssymb,amsmath}
\usepackage{graphicx}
\usepackage{bm}
\usepackage{caption,subcaption}
\usepackage{tikz}
\usepackage{soul}
\usepackage{listings}

\usepackage[normalem]{ulem}

\renewcommand\Re{\mathrm{Re\,}}
\renewcommand\Im{\mathrm{Im\,}}

\def\be{\begin{equation}}
\def\ee{\end{equation}}
\def\bea{\begin{eqnarray}}
\def\eea{\end{eqnarray}}
\def\e{\epsilon}

\usepackage{xcolor}

\graphicspath{./figures/}

\newsavebox{\twosubbox}

\begin{document}

\title{Spatially Extended Dislocations Produced by the Dispersive Swift-Hohenberg Equation}

\author{Brenden Balch}
\affiliation{Department of Mathematics, Colorado State University, Fort
Collins, CO 80523, USA}

\author{Patrick D. Shipman}
\affiliation{Department of Mathematics and School of Advanced Materials Discovery, Colorado State University, Fort
Collins, CO 80523, USA}

\author{R. Mark Bradley}
\affiliation{Departments of Physics and Mathematics, Colorado State University, Fort
Collins, CO 80523, USA}

\date{\today}

\begin{abstract}

Motivated by previous results showing that the addition of a linear dispersive term to the two-dimensional Kuramoto-Sivashinsky equation has a dramatic effect on the pattern formation, we study the Swift-Hohenberg equation with an added linear dispersive term, the dispersive Swift-Hohenberg equation (DSHE).  
The DSHE produces stripe patterns with spatially extended dislocations that we call seam defects. In contrast to the dispersive Kuramoto-Sivashinsky equation, the DSHE has a narrow band of unstable wavelengths close to an instability threshold.  This allows for analytical progress to be made.  We show that the amplitude equation for the DSHE close to threshold is a special case of the anisotropic complex Ginzburg-Landau equation (ACGLE) and that seams in the DSHE correspond to spiral waves in the ACGLE.  Seam defects and the corresponding spiral waves tend to organize themselves into chains, and we obtain formulas for the velocity of the spiral wave cores and for the spacing between them.   In the limit of strong dispersion, a perturbative analysis yields a relationship between the amplitude and wavelength of a stripe pattern and its propagation velocity.  Numerical integrations of the ACGLE and the DSHE confirm these analytical results.

\end{abstract}

\maketitle

\vfill\eject

\section{Introduction}
\label{sec: intro}

The Kuramoto-Sivashinsky (KS) equation occurs in many contexts, including the nonlinear evolution of flame fronts \cite{Sivashinsky1983}, concentration waves in reaction-diffusion systems \cite{Kuramoto1976}, and nanoscale pattern formation produced by bombardment of a solid surface with a broad ion beam \cite{Cuerno1995,Makeev2002,Bradley2020}. It is among the simplest partial differential equations that exhibit spatiotemporal chaos. Adding a linearly dispersive term to the one-dimensional (1D) KS equation yields the dispersive KS equation in 1D,
\be 
u_t=-u_{xx}-u_{xxxx}+u_x^2 + \gamma u_{xxx},
\label{1DKS}
\ee 
where \(u=u(x,t)\) and \(\gamma\) is real. (The 1D KS equation is recovered for \(\gamma=0\)).  Surprisingly, when \(\gamma\) is large and the initial condition is low amplitude spatial white noise, highly ordered patterns emerge at sufficiently long times and the spatio-temporal chaos that would otherwise prevail is suppressed \cite{Kawahara1983}.  This remains true if a strong linearly dispersive term is added to the anisotropic KS equation in two dimensions (2D) \cite{Loew2019,Bradley2020}.  

In the limit that \(\gamma\) tends to infinity, the 1D dispersive KS equation (\ref{1DKS}) becomes the Korteweg-DeVries (KdV) equation.  The KdV equation has solutions in which multiple solitons are present.  For large but finite \(\gamma\), there is a repulsive interaction between neighboring solitons, and the solitons eventually arrange themselves in an ordered chain as a consequence \cite{Kawahara1988}.  Thus, there is some understanding of how order emerges in solutions of Eq.~(\ref{1DKS}) for \(\gamma\gg 1\).  This picture does not carry over to the anisotropic 2D KS equation with added dispersion, however.  

When a solid surface is bombarded with a broad ion beam and the angle of ion incidence \(\theta\) exceeds a threshold value \(\theta_c\), self-assembled ripples with wavelengths as short as 10 nm form \cite{Munoz-Garcia2014}.  If the patterns formed were not almost always disordered, ion bombardment could become a widely employed method of fabricating large-area nanostructures with feature sizes too small to be attained by conventional optical lithography.  After rescaling, the equation that describes the time evolution of an ion-bombarded solid surface for \(\theta\) just above \(\theta_c\) is
\be 
u_t=-u_{xx}-u_{xxxx}+u_x^2 + u_{yy} + \gamma u_{xxx},
\label{eom cr}
\ee 
where \(u=u(x,y,t)\) is the height of the solid surface about the point \((x,y)\) in the \(x-y\) plane at time \(t\) and \(\gamma\propto (\theta - \theta_c)^{-1/2}\) diverges as \(\theta\to\theta_c^+\) \cite{Bradley2020}.  Equation (\ref{eom cr}) reduces to Eq.~(\ref{1DKS}) if \(u\) is independent of \(y\). It is a simplified version of the anisotropic 2D KS equation with linear dispersion, and simulations show that it produces highly ordered ripples if \(\gamma\) is large, i.e., if \(\theta\) is just above \(\theta_c\) \cite{Loew2019,Bradley2020}.  This finding has the potential to revolutionize the field of nanoscale patterning by ion bombardment, and, accordingly, it is of considerable importance to understand how strong linear dispersion modifies the dynamics.  

A second intriguing observation emerges from simulations of Eq.~(\ref{eom cr}):  dispersion can lead to the formation of transient raised and depressed triangular regions that are traversed by ripples for moderate values of \(\gamma\).  Triangular nanostructures of this kind have been observed in many experiments in which a solid surface is bombarded with an obliquely incident ion beam \cite{Munoz-Garcia2014,Carter1977,Keller2008,Keller2008b,Metya2012,Chowdhury2013,Teichmann2013,Teichmann2014,Chowdhury2016,Gago2018,Lopez-Cazalilla2018}, but their formation is currently poorly understood. In simulations, once the triangular nanostructures have disappeared, the surface has a disordered appearance with streaks parallel to the \(x\) axis.

The Swift-Hohenberg equation (SHE) is an important model equation in the study of pattern formation in spatially extended nonlinear systems \cite{Cross2009}.  Close to the threshold for pattern formation, analytical results can be obtained because there is a narrow band of unstable wavelengths.  In particular, the amplitude equation, which describes the slow variation of the pattern in space and time, can be derived. 

In this paper, we study the SHE with added linear dispersion in both one and two dimensions. Our motivation for doing so is this: the effect of strong linear dispersion can be better understood in the context of the SHE than for the KS equation because there is a narrow band of unstable wavelengths close to threshold in the case of the SHE.  We find that the 2D dispersive Swift-Hohenberg equation (DSHE) produces a unique type of spatially extended defect if the linear dispersion is sufficiently strong.  These defects --- which we will refer to as \lq\lq seams'' --- are essentially dislocations that are smeared out along a line segment oriented obliquely to the \(x\) axis.  As we will discuss, these are related to the triangular nanostructures that are observed when a solid surface is bombarded with a broad ion beam.

Simplicity emerges in the DSHE in two limits: close to threshold and in the limit of strong dispersion.   
Close to threshold, we show that the amplitude equation for the DSHE is a special case of the anisotropic complex Ginzburg-Landau equation (ACGLE).  The seams in the original equation of motion are spiral waves in the ACGLE.  These spiral waves and the corresponding seam defects tend to arrange themselves into chains.   We predict the velocity of the spiral wave cores and the spacing between them for a particular type of controlled initial condition.
In the limit of strong dispersion, on the other hand, we carry out a perturbative analysis that shows that the stripes have a nearly sinusoidal dependence on position. The analysis also yields the stripe's propagation velocity and a relationship between their amplitude and wavelength.  These predictions are in excellent accord with the results of our numerical integrations of the equation of motion.

This paper is organized as follows: In Sec.~\ref{sec:DSHE}, we recast the DSHE in dimensionless form and perform a linear stability analysis. We find an approximate solution to the 1D DSHE in the limit of strong linear dispersion in Sec.~\ref{sec:1dpert}. In Sec.~\ref{sec:amp}, we derive the amplitude equation that applies close to the threshold for pattern formation.  Simulations of the DSHE and the corresponding amplitude equation are carried out in Sec.~\ref{sec:numericalsimulations}.  We also study the dynamics of chains of spiral waves both analytically and numerically. Our work is placed in context in Sec.~\ref{sec:discussion}, and we conclude in Sec.~\ref{sec:conclusions}.

\section{The Dispersive Swift-Hohenberg Equation}
\label{sec:DSHE}

In this paper, we study the DSHE
\begin{align}
    u_t = -a \Delta^2 u - b u_{xx} +  c u_{yy} + d u_{xxx} + e u - f u^3
    \label{init eom}
\end{align}
in one and two dimensions.  Here \(u=u(x,y,t)\) and \(a\), \(b,\ldots\), and  \(f\) are real parameters. We confine our attention to the case in which \(a\), \(b\) and \(f\) are positive. We introduce the dimensionless parameters \(\tilde{u} = 2(a f/b^2)^{1/2}u\), \(\tilde{x} = \mathrm{sgn}(d)[b/(2a)]^{1/2}x\), \(\tilde{y} = \mathrm{sgn}(d)[b/(2a)]^{1/2}y\), and \(\tilde{t} = [b^2/(4a)]t\). Dropping the tildes, we find the rescaled equation of motion to be
\begin{align}
\label{eom2d}
    u_t = -\Delta^2u - 2(u_{xx} - \beta u_{yy}) + \gamma u_{xxx} + (\mu -1) u - u^3,
\end{align}
where \(\mu = 1+4ae/b^2\), \(\beta = c/b\) and \(\gamma = [2 d^2/(a b)]^{1/2}\). Note that \(\gamma\) is nonnegative. For the case \(\gamma=0\), there is no dispersion and Eq.~(\ref{eom2d}) reduces to the usual SHE.

The equation of motion (\ref{eom2d}) has the equilibrium solution \(u=0\). Linearizing about this solution, we obtain
\begin{align}
    u_t = -\Delta^2u - 2(u_{xx} - \beta u_{yy}) + \gamma u_{xxx} + (\mu -1) u.
\end{align}
Setting \(u = \exp({i \vec{k}\cdot \vec{x}+\sigma t})\), we find the dispersion relation
\begin{align}
    \sigma = -k^4 + 2(k_x^2 - \beta k_y^2) + \mu -1 -i\gamma k_x^3,
\end{align}
where \(\vec{k} = (k_x,k_y)\) is the wave vector. An easy calculation shows that \(\Re{\sigma}\) is maximized for \(\vec{k}=(\pm1,0)\) and has the maximum value \(\mu\) provided that \(\beta>-1\), which we assume to be the case. This tells us that the solution \(u=0\) is linearly stable when \(\mu<0\) and linearly unstable whenever \(\mu>0\). By the continuity of \(\Re{\sigma}=\Re\sigma(\vec k)\), it follows that there are neighborhoods about the points \(\vec{k} = (\pm1,0)\) in which \(\Re{\sigma}\) is positive if \(\mu>0\). For small, positive \(\mu\), neither neighborhood contains the zero vector, indicating a type-I instability. Moreover, the phase velocity is
\begin{align}
    v = -\frac{\Im{\sigma}}{k} = \gamma \frac{k_x^3}{k}.
    \label{lin phase vel}
\end{align}
For the 1D case in which \(u_y=0\), the phase velocity (\ref{lin phase vel}) reduces to \(v= \gamma k_x^2\).

\section{The Strongly Dispersive Limit}
\label{sec:1dpert}
We begin by studying the equation of motion (\ref{eom2d}) when dispersion is strong, i.e., the case in which \(\gamma\gg 1\).  We set \(u=u(x,t)\) in Eq.~(\ref{eom2d}) and so obtain
\begin{align}
    \label{eom1d}
    u_t = -(1+\partial_x^2)^2u + \gamma u_{xxx} + \mu u - u^3.
\end{align}
We seek solutions to Eq.~(\ref{eom1d}) of the form \(u=u(x-vt)\). Moreover, we will set \(\epsilon = \gamma^{-1}\) and take \(\gamma\) to be large. Equation~(\ref{eom1d}) now yields 
\begin{align}
    \label{1dpert}
    u_{xxx} + \omega u_x + \epsilon\left[\mu u-(1+\partial_x^2)^2u-u^3\right]=0,
\end{align}
where \(\omega \equiv v/\gamma = \epsilon v\). Next, we assume that
\begin{align}
    u &= u_0 + \epsilon u_1 + \epsilon^2 u_2 + \mathrm{h.o.t.} ~~ \text{and}\\
    \omega &= \omega_0 + \epsilon \omega_1 +\epsilon^2 \omega_2 + \mathrm{h.o.t.,}
\end{align}
where h.o.t.~stands for higher-order terms.  Then, to zeroth order in \(\e\), Eq.~(\ref{1dpert}) reads
\begin{align}
    \label{zeroth}
    u_{0xxx} + \omega_0 u_{0x} = 0.
\end{align}
The general solution to Eq.~(\ref{zeroth}) is given by
\begin{align}
    u_0 = C+A\cos(\sqrt{\omega_0} x + \phi),
\end{align}
where \(C\), \(A\) and \(\phi\) are arbitrary constants.
By choosing the origin appropriately, we may arrange for \(\phi\) to be zero. Thus, we have
\begin{align}
    \label{gen0}
    u_0 = C+A\cos(kx),
\end{align}
where \(k\equiv\sqrt{\omega_0}\) is the wave number. 

To first order, Eq.~(\ref{1dpert}) may be written
\begin{align}
    \label{first}
    u_{1xxx}+\omega_0 u_{1x} &= -\omega_1 u_{0x} -\mu u_0 + (1+\partial_x^2)^2u_0 + u_0^3\nonumber\\
    &\equiv q.
\end{align}
Let \(L_0 = \partial_x^3 + k^2 \partial_x\), so that Eq.~(\ref{first}) can be written compactly as 
\begin{align}
    L_0u_1=q.
\end{align}
It is a straightforward exercise to show that \(L_0:C^3[-L,L]\to C[-L,L]\) is a Fredholm operator. The Fredholm Alternative then implies that \(q\) is orthogonal to \(\ker L_0^\dagger\), where \(L_0^\dagger\) denotes the adjoint with respect to the \(L^2\) inner product. Because
\begin{align}
    \ker L_0^\dagger = \text{span}\{1,e^{ikx},e^{-ikx}\},
\end{align}
the constant term in \(q\) must be zero, i.e.,
\begin{align}
    -\mu C+C+C^3=0.
\end{align}
Equivalently, \(C=0\) or \(C^2=\mu-1\). Since \(C\) is real, the latter possibility is ruled out whenever \(\mu<1\), and we take this to be the case. This means that
\begin{align}
    u_0 = A\cos(kx).
\end{align}
Further still, we have
\begin{align}
    q&=\omega_1 k A\sin(kx) -\mu A\cos(kx) + A(1-k^2)^2\cos(kx) + A^3\cos^3(kx)\\
    &=\omega_1 k A\sin(kx) -\mu A\cos(kx) + A(1-k^2)^2\cos(kx) + \frac14A^3\left[\cos(3kx)+3\cos(kx)\right].
\end{align}
\(q\in(\ker L_0^\dagger)^\perp\) therefore implies that \(\omega_1=0\) and
\begin{align}
    \label{pertamp}
    A^2=\frac43 \left[\mu-(1-k^2)^2\right].
\end{align}
Because \(A^2\geq 0\), we must have
\begin{align}
    \label{band}
    \left|1-k^2\right|\leq\sqrt{\mu}.
\end{align}
This establishes that a steady-state, propagating solution is obtained only for wave numbers in the linearly unstable band.  We also see that \(A^2 = 4\Re\sigma(\vec{k})/3\), and so we come to the natural conclusion that the higher the linear growth rate, the higher the amplitude of the corresponding steady-state solution.
Now note that Eq.~(\ref{first}) reduces to
\begin{align}
    \label{redfirst}
    u_{1xxx}+k^2u_{1x}= \frac14 A^3\cos(3kx).
\end{align}
We will seek a solution to Eq.~(\ref{redfirst}) of the form
\begin{align}
    u_1=B\sin(3kx).
\end{align}
In doing so, we obtain
\begin{align}
    B=\frac{1}{96}k^{-3}A^3,
\end{align}
and hence
\begin{align}
    u(x,t) = A\cos(k(x-vt)) + \frac{1}{96}k^{-3}A^3\epsilon\sin(3k(x-vt)) + O(\epsilon^2),
    \label{u with correction}
\end{align}
where \(A\) and \(k\) satisfy Eqs.~(\ref{pertamp}) and (\ref{band}), respectively.
Equation (\ref{u with correction}) gives the approximate form of the propagating, periodic solution to Eq.~(\ref{eom1d}). The presence of the correction with wave number \(3k\) in Eq.~(\ref{u with correction}) is to be expected because a cubic nonlinearity is present in the equation of motion (\ref{eom1d}).

Since \(\omega_0 = k^2\) and \(\omega_1=0\),
\begin{align}
    \omega = \omega_0 + \e\omega_1 + O(\epsilon^2) = k^2 + O(\epsilon^2).
\end{align}
This in turn gives us the phase velocity,
\begin{align}
    \label{pertvel}
    v = \gamma k^2 + O(\epsilon).
\end{align}
This shows that in the strongly dispersive (\(\gamma\to\infty\)) limit, the phase velocity (\ref{pertvel}) obtained by a perturbative analysis of the full nonlinear equation of motion reduces to the phase velocity (\ref{lin phase vel}) for the linearized problem.

If we begin a simulation of the equation of motion (\ref{eom1d}) with a low amplitude spatial white noise initial condition, it is not evident whether the solution will evolve toward a solution of the form (\ref{u with correction}) with the phase velocity given by Eq.~(\ref{pertvel}) and with \(A\) and \(k\) related by Eq.~(\ref{pertamp}).  Even if that turns out to be the case, it is not clear {\it a priori} what the chosen value of \(k\) will be, although the inequality (\ref{band}) would have to be satisfied.  Numerical integrations of Eq.~(\ref{eom1d}) will be carried out in Section~\ref{sec:numericalsimulations} to address these issues.

\section{Near-Threshold Behavior}
\label{sec:amp}

In this section, we analyze the equation of motion (\ref{eom2d}) close to threshold, i.e., for small, positive \(\mu\).  Because we have assumed that \(\beta>-1\), there are small neighborhoods about the critical wave vectors \(\vec{k} = (\pm1,0)\) in which \(\Re\sigma(\vec k)\) is positive. 
This implies the existence of an amplitude equation. To find this amplitude equation, we begin by writing Eq.~(\ref{eom2d}) as
\be 
u_t = \mathcal{L} u - u^3,
\label{eom2d2}
\ee 
where
\be 
\mathcal{L}\equiv -\Delta^2 -2(\partial_x^2-\beta\partial_y^2)+\gamma\partial_x^3 + \mu -1
\ee 
is the linear part of the differential operator on the right-hand side of Eq.~(\ref{eom2d}).
The linear dispersion relation tells us that, to leading order, the solution to Eq.~(\ref{eom2d}) is a traveling plane wave with wave number \(k=1\) that propagates in the \(x\) direction. Note that the phase velocity of the mode with wave vector \(\vec{k} = (1,0)\) is \(\gamma\), and the corresponding group velocity is \(3\gamma\). Accordingly, we begin with the ansatz
\begin{align}
    \label{ansatz}
    u &= \mu^{1/2}u_0 + \mu u_1 + \mathrm{h.o.t.}\nonumber\\
    &= \mu^{1/2} A(\xi,Y,T)e^{i(x-\gamma t)}+\mathrm{c.c.} + \mu u_1 + \mathrm{h.o.t.,}
\end{align}
where \(\xi \equiv \mu^{1/2}(x- 3 \gamma t)\), \(Y\equiv\mu^{1/2}y\) and \(T \equiv \mu t\) are slow variables and c.c.~denotes the complex conjugate.
As a result, we must make the replacements
\(\partial_x \mapsto \partial_x + \mu^{1/2} \partial_\xi\), \(\partial_y \mapsto \mu^{1/2} \partial_Y\) and \(\partial_t \mapsto \partial_t -  3\mu^{1/2} \gamma \partial_\xi + \mu \partial_T\) in Eq.~(\ref{eom2d2}).  This leads to
\be 
\mathcal{L}\mapsto L_0 + \mu^{1/2} L_1 + \mu L_2+\mathrm{h.o.t.,}
\ee 
where
\begin{align}
    L_0 &= \gamma\partial_x^3 - (\partial_x^2+1)^2\\
    L_1 &= \left(-4\partial_x^3 + 3\gamma\partial_x^2 - 4 \partial_x\right)\partial_\xi\\
    L_2 &= -6 \partial_\xi^2 \partial_x^2 -2 \partial_x^2 \partial_Y^2 + 3\gamma \partial_\xi^2 \partial_x -2 \partial_\xi^2 +2\beta \partial_Y^2 +1.
\end{align}

To order \(\mu^{1/2}\), Eq.~(\ref{eom2d2}) is
\begin{align}
    \partial_tu_0=L_0u_0.
    \label{lowest order}
\end{align}
This automatically holds since we set 
\be 
u_0 = A(\xi,Y,T)e^{i(x-\gamma t)}+\mathrm{c.c.}
\label{u0}
\ee

To order \(\mu\), Eq.~(\ref{eom2d2}) yields
\begin{align}
    \partial_t u_1 - 3 \gamma \partial_\xi u_0 = L_1 u_0 + L_0 u_1.
    \label{foo}
\end{align}
Since \(L_1 u_0 = - 3 \gamma \partial_\xi u_0\), Eq.~(\ref{foo}) reduces to
\begin{align}
    \partial_t u_1= L_0u_1. 
\end{align}
This merely tells us that
\begin{align}
    u_1 = A_1(\xi,Y,T)e^{i(x-\gamma t)} +\mathrm{c.c.}
    \label{u1}
\end{align}

To order \(\mu^{3/2}\), Eq.~(\ref{eom2d2}) gives
\begin{align}
    \partial_T u_0 - 3 \gamma \partial_\xi u_1 + \partial_t u_2 = L_2 u_0 + L_1 u_1 + L_0 u_2 - u_0^3.
    \label{3rd order}
\end{align}
Next, using Eq.~(\ref{u0}), Eq.~(\ref{3rd order}) can be rearranged to obtain 
\begin{align}
    \Lambda u_2 = & \left[-A_T + A + 4\left(1+i\frac34 \gamma\right)A_{\xi\xi} + 2\left(1+\beta\right)A_{YY} - 3\left|A\right|^2 A\right]e^{i(x-\gamma t)}\nonumber\\
    &- A^3 e^{3i(x-\gamma t)} + \mathrm{c.c.},\nonumber\\
    \equiv &\,\, Q,
    \label{finalorder}
\end{align}
where \(\Lambda \equiv \partial_t - \gamma\partial_x^3 + (\partial_x^2+1)^2\). A quick check shows that \(e^{i(x-\gamma t)} \in \ker \Lambda\), which implies that
\begin{align}
    \label{coframe}
    A_T = A + 4\left(1+ i\frac34 \gamma\right)A_{\xi\xi} + 2\left(1+\beta\right)A_{YY} - 3\left|A\right|^2 A.
\end{align}
Equation (\ref{coframe}) is the amplitude equation for the two-dimensional (2D) DSHE, Eq.~(\ref{eom2d}).
If we put \(A = \tilde A/\sqrt{\mu}\) in Eq.~(\ref{coframe}), drop the tilde, and write the result in terms of the original coordinates, we obtain
\begin{align}
    \label{amp2d}
    A_t + 3\gamma A_x = \mu A + 4\left(1+ i\frac34 \gamma \right)A_{xx}+ 2(1+\beta)A_{yy} - 3\left|A\right|^2 A.
\end{align}
We prefer, however, to put the amplitude equation (\ref{coframe}) in the standard form used in Refs.~\cite{handwerk2020} and \cite{faller99} by setting \(\hat A=\sqrt{3}A\),
\(\hat x=\xi/2\), \(\hat y=Y/\sqrt{2(1+\beta)}\) and \(\hat t = T\) and then dropping the hats.  This gives
\begin{align}
    \label{standardamp}
    A_t = A + (1+i\eta)A_{xx} + A_{yy} - |A|^2 A,
\end{align} 
where \(\eta\equiv 3\gamma/4\). Equation~(\ref{standardamp}) is a special case of the ACGLE \cite{faller99,handwerk2020}. If there is no dispersion, then \(\gamma=0\) and Eq.~(\ref{standardamp}) reduces to the isotropic (real) Ginzburg-Landau equation. 

\section{Numerical simulations}
\label{sec:numericalsimulations}
We carry out numerical simulations of Eq.~(\ref{eom1d}) on \(x\in[-L,L]\), and of Eqs.~(\ref{eom2d}) and (\ref{standardamp}) on the square domain \((x,y)\in[-L,L]^2\). To do so, we employ Fourier spectral methods with periodic boundary conditions, coupled with the fourth order exponential time differencing Runge-Kutta method (ETDRK4). Implementations of this method can be found in Refs.~\cite{kassamspec} and \cite{crasterspec}, while full derivations of the method can be found in Refs.~\cite{cox2002exponential} and \cite{krogstad05}. In all simulations in this paper, we employ a spatial grid with \(N=2048\) grid points in 1D and an \(N \times N\) spatial grid with \(N=128\) in the 2D simulations unless otherwise noted.  The time step in all cases is \(\Delta t=0.01\).

\subsection{Simulations of the Dispersive Swift-Hohenberg Equation}
\label{sec:shsims}

\begin{figure}
    \centering
    \includegraphics[width=\textwidth]{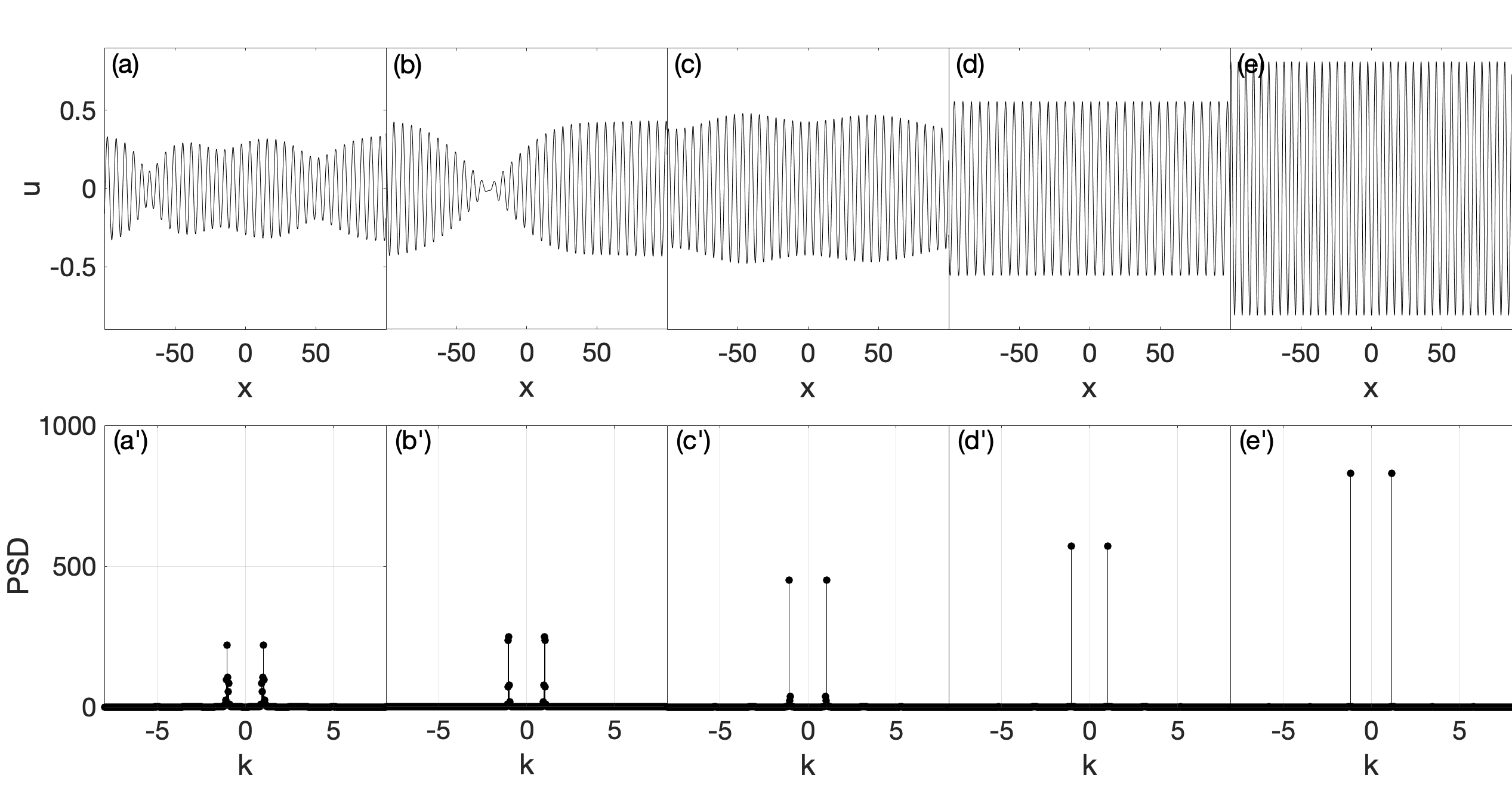}
    \caption{The first row depicts solutions to Eq.~(\ref{eom1d}) on the spatial domain \(x\in [-100,100]\), and the second row shows the corresponding PSDs. In all cases, \(\mu = 0.1\), which is near the threshold for pattern formation. From left to right, \(\gamma =0\), 25, 50, 75 and 100. All images are for time \(t=100\).}
    \label{fig:1dsols}
\end{figure}

Figure~\ref{fig:1dsols} shows results of simulations of the 1D equation of motion Eq.~(\ref{eom1d}) and the corresponding power spectral densities (PSDs) at time \(t=100\) for \(\mu=0.1\) and selected values of \(\gamma\). The initial conditions were low amplitude spatial white noise. The simulations suggest that as \(\gamma\) gets large, the solution tends to a sinusoidal form, in accord with the perturbation theory prediction.

\begin{figure}
    \centering
    \includegraphics[width=\textwidth]{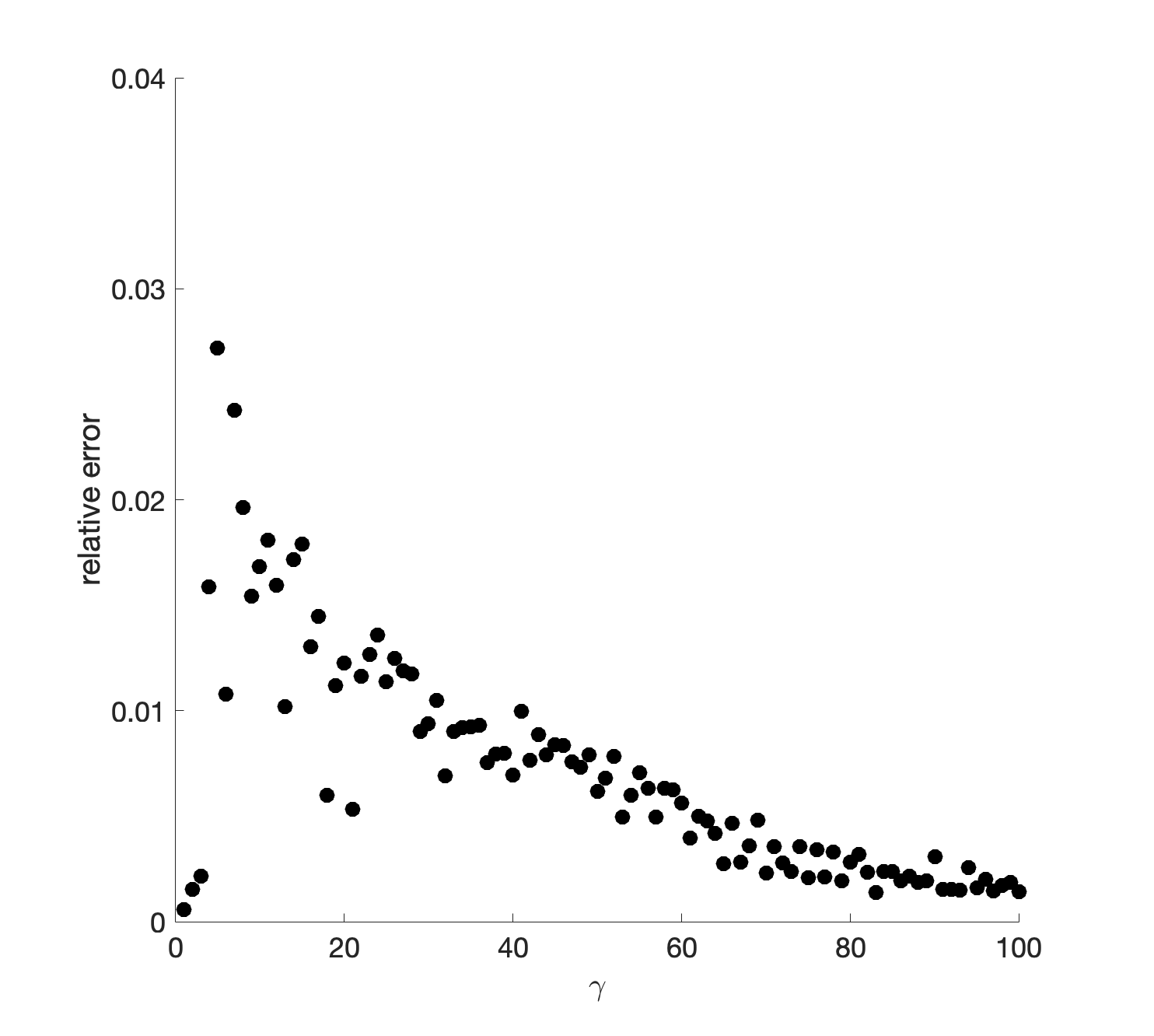}
    \caption{Comparison of Eq.~(\ref{pertamp}) to simulation results for values of \(\gamma\) between \(1\) and \(100\). Each data point gives the relative error of the amplitude for the corresponding value of \(\gamma\). In each simulation, \(\mu=1\), the domain was \(x\in [-100,100]\), and the measurements were taken at \(t=100\). We note that the relative error is less than \(1\%\) for values of \(\gamma\) larger than 50, and decreases as \(\gamma\) increases.}
    \label{fig:1damp_pert}
\end{figure}

\begin{figure}
    \centering
    \includegraphics[width=\textwidth]{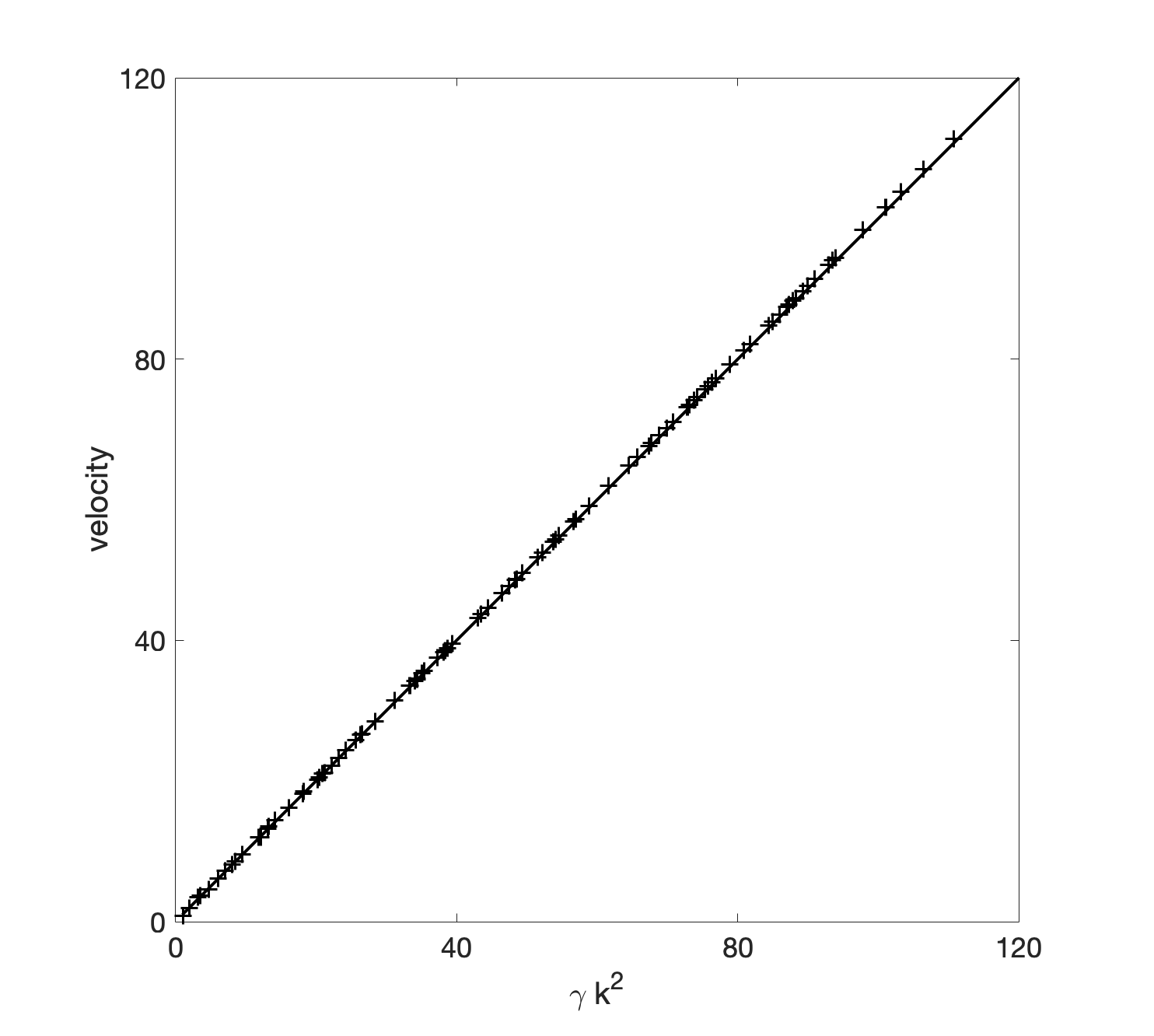}
    \caption{The phase velocity of the steady-state propagating solution versus \(\gamma k^2\), as computed from numerical simulations (+'s). Each point is the result of a single simulation with \(\mu=1\) and a value of \(\gamma\) between 1 and 100. The solid line shows the theoretical prediction. The domain for each simulation was \(x\in [-100,100]\), and the velocities were calculated at time \(t=100\).}
    \label{fig:vel1d}
\end{figure}

The perturbation theory prediction (\ref{pertamp}) gives the amplitude as a function of the wave number \(k\) to order \(\gamma^{-1}\).  Figure~\ref{fig:1damp_pert} shows the relative error in Eq.~(\ref{pertamp}), where the relative error is defined to be the absolute value of the difference between the measured and predicted values divided by their sum. Note that as \(\gamma\) increases, the relative error decreases and is less than \(1\%\) when \(\gamma>50\). Thus, Eq.~(\ref{pertamp}) appears to hold in the limit \(\gamma\to \infty\), as expected. 
The perturbation theory also predicts the phase velocity of the solution.  In the simulations, the observed velocity was taken to be \(\Delta \phi/(k \Delta t)\), where \(\Delta \phi\) is the phase difference in \(u\) at two times separated by time \(\Delta t\)  and \(k\) is the dominant wave number.  Figure~\ref{fig:vel1d} compares the prediction given by Eq.~(\ref{pertvel}) to the observed velocities determined from 100 simulations --- one for each integer value of \(\gamma\) between zero and 99. The simulations were run until time \(t=100\) and the velocities were determined from the last two time steps.  Figure~\ref{fig:vel1d} is another indication that the simulated results agree very well with perturbation theory.

\begin{figure}
    \centering
    \includegraphics[width=\textwidth]{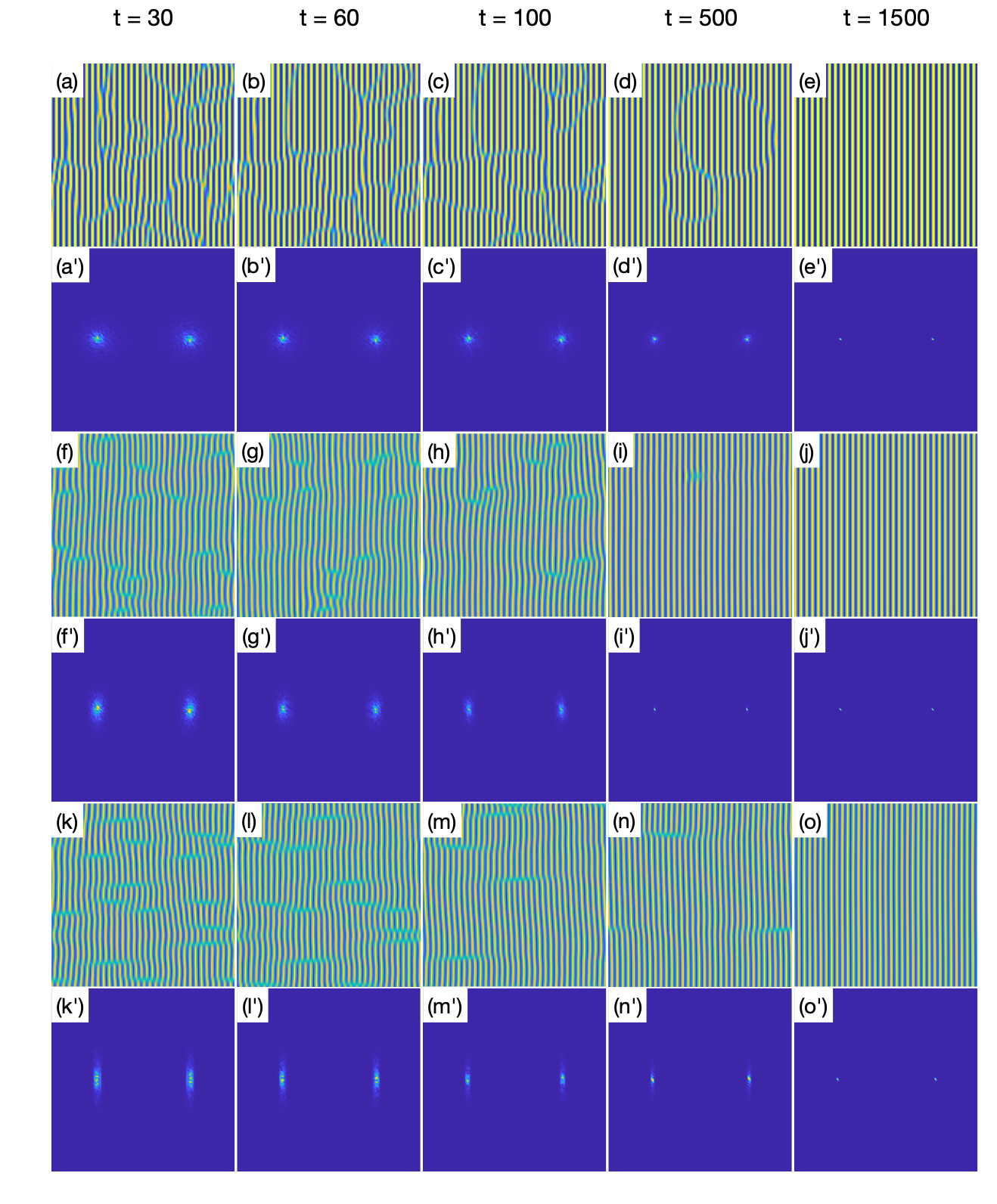}
    \caption{Solutions to Eq.~(\ref{eom2d}) with parameters \(\mu=\beta=1\) on the domain \((x,y)\in [-100,100]^2\).  The values of \(\gamma\) are the \(0,10\) and \(100\) for the first, second and third pairs of rows, respectively. In each pair of rows, the first row shows the solution at the times listed and the second row shows the corresponding PSDs.  The columns from left to right depict the solutions at times \(t=30, 60, 100, 500\) and \(1500\).}
    \label{fig:shw2d}
\end{figure}

\begin{figure}
    \centering
    \includegraphics[width=\textwidth]{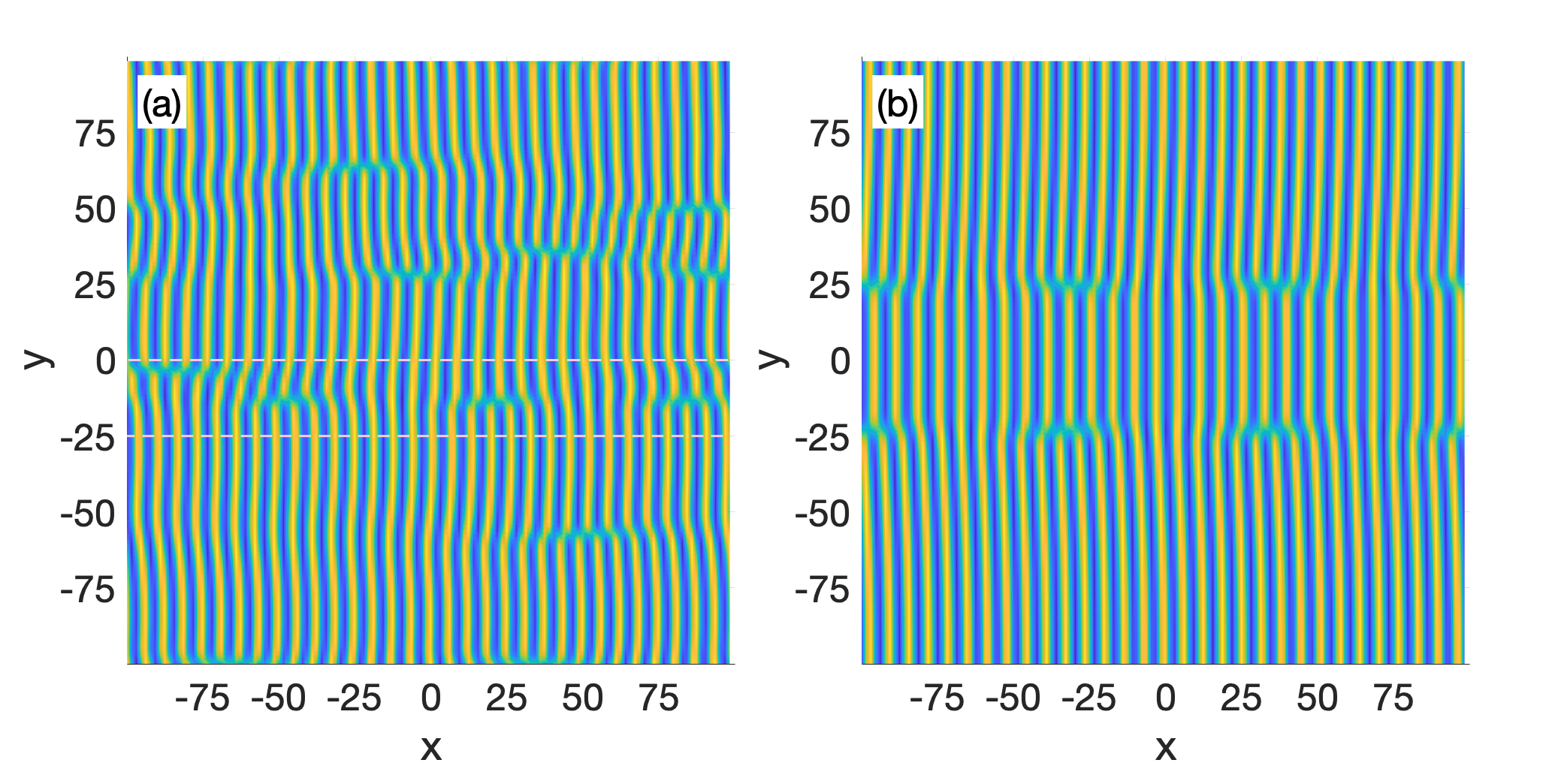}
    \caption{(a) A solution to Eq.~(\ref{eom2d}) at time \(t=40\) that was started with a low amplitude spatial white noise initial condition. Note the chain of three defects between the horizontal lines. (b) A solution to Eq.~(\ref{eom2d}) with a banded initial condition of the form (\ref{eomchainsIC}) at time \(t=100\). The initial condition had \(n_1=28\) and \(n_2=31\).   The parameter values were \(\mu=\beta=1\) and \(\gamma=100\) in both (a) and (b).}
    \label{fig:eomchains}
\end{figure}

Turning our attention to the 2D case, Fig.~\ref{fig:shw2d} shows the time evolution of solutions to Eq.~(\ref{eom2d}) and their corresponding PSDs for \(\gamma=0\), 10 and 100. In each case, \(\mu=1\). For \(\gamma=0\), Eq.~(\ref{eom2d}) is just the Swift-Hohenberg equation. We included a simulation for this case for reference. For \(\gamma > 0\), the defects are stretched dislocations or seams which are obliquely oriented with respect to the \(x\) axis. The phase changes through \(\pm 2\pi\) on a contour that circles a seam.  Of particular note is the appearance of several seams at nearly the same \(y\) value but differing values of \(x\). We call these defect chains. Figure~\ref{fig:eomchains} (a) shows a solution to Eq.~(\ref{eom2d}) for a spatial white noise initial condition with a chain of three defects. These chains of seam defects are present at early times. At later times, defects of opposite sign meet and mutually annihilate, which ultimately results in a defect-free pattern. 

\begin{figure}
    \centering
    \includegraphics[width=\textwidth]{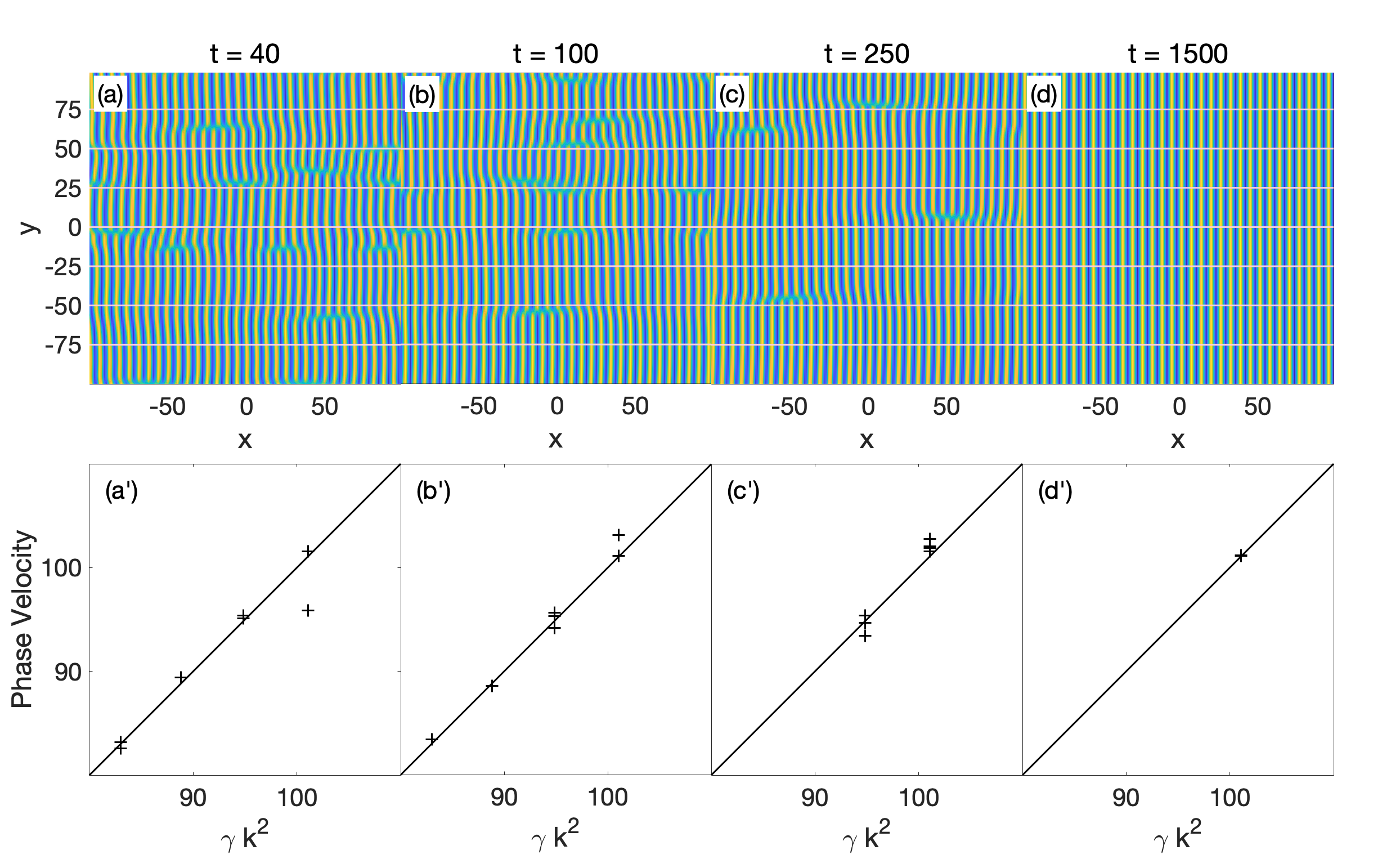}
    \caption{The time evolution of a solution to Eq.~(\ref{eom2d}) with \(\mu=\beta=1\) and \(\gamma=100\) is shown in the first row. The domain is \((x,y)\in [-100,100]^2\). The phase velocity in the \(x\) direction was computed for each of the cuts parallel to the \(x\) axis that are shown. The second row shows the observed velocities along each cut (+'s) versus the velocities predicted by Eq.~(\ref{pertvel}) (solid lines). }
    \label{fig:vel2d}
\end{figure}

The time evolution that occurs with relatively large \(\gamma\) in one and two dimensions is similar in several ways. In 2D, after some time, multiple roughly horizontal bands have formed in which \(u\) is almost independent of \(y\), as seen in Fig.~\ref{fig:eomchains} (a).  These bands are separated by chains of seam defects. Within a band, the form of the solution is close to a solution to the 1D DSHE, and so the the phase velocity is approximately equal to \(\gamma k^2\). Figure \ref{fig:vel2d} shows the time evolution of a solution. Defects are present except at the latest time, \(t=1500\). For each of the cuts parallel to the \(x\) axis that are shown in Fig.~\ref{fig:vel2d} (a) - (d), the velocity in the \(x\) direction was computed and compared to Eq.~(\ref{pertvel}).  The results of this comparison are shown in Fig.~\ref{fig:vel2d} (a') - (d'). The agreement is very good at each of the four times shown in the figure, except where a cut passes directly through a seam.

\begin{figure}
    \centering
    \includegraphics[width=\textwidth]{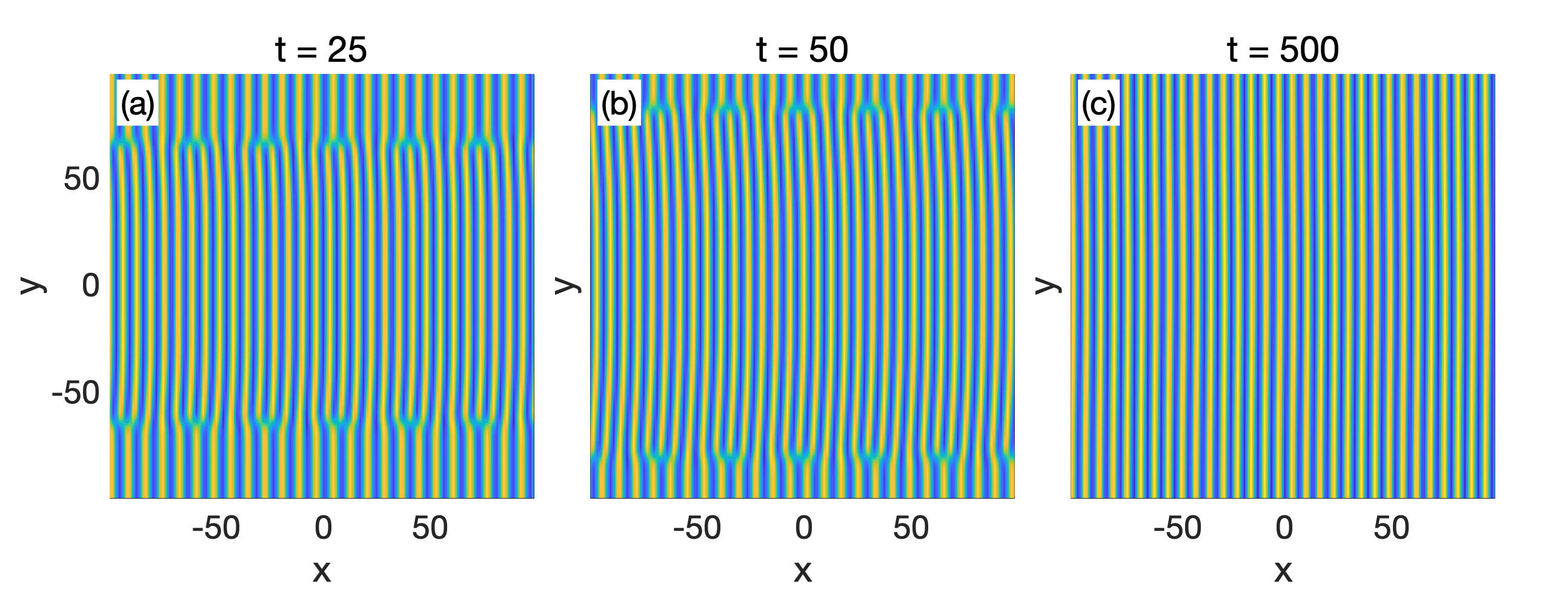}
    \caption{A solution to Eq.~(\ref{eom2d}) with the parameter values \(\mu=\beta=1\) and \(\gamma = 100\) on the domain \((x,y)\in [-100,100]^2\) at times (a) \(t=25\), (b) \(t=50\) and (c) \(t=500\). The initial condition was given by Eq.~(\ref{eomchainsIC}) with \(k_1\) and \(k_2\) chosen so that \(n_1=25\) and \(n_2=31\).}
    \label{fig:chain_annihilation}
\end{figure}

\begin{figure}
    \centering
    \includegraphics[width=\textwidth]{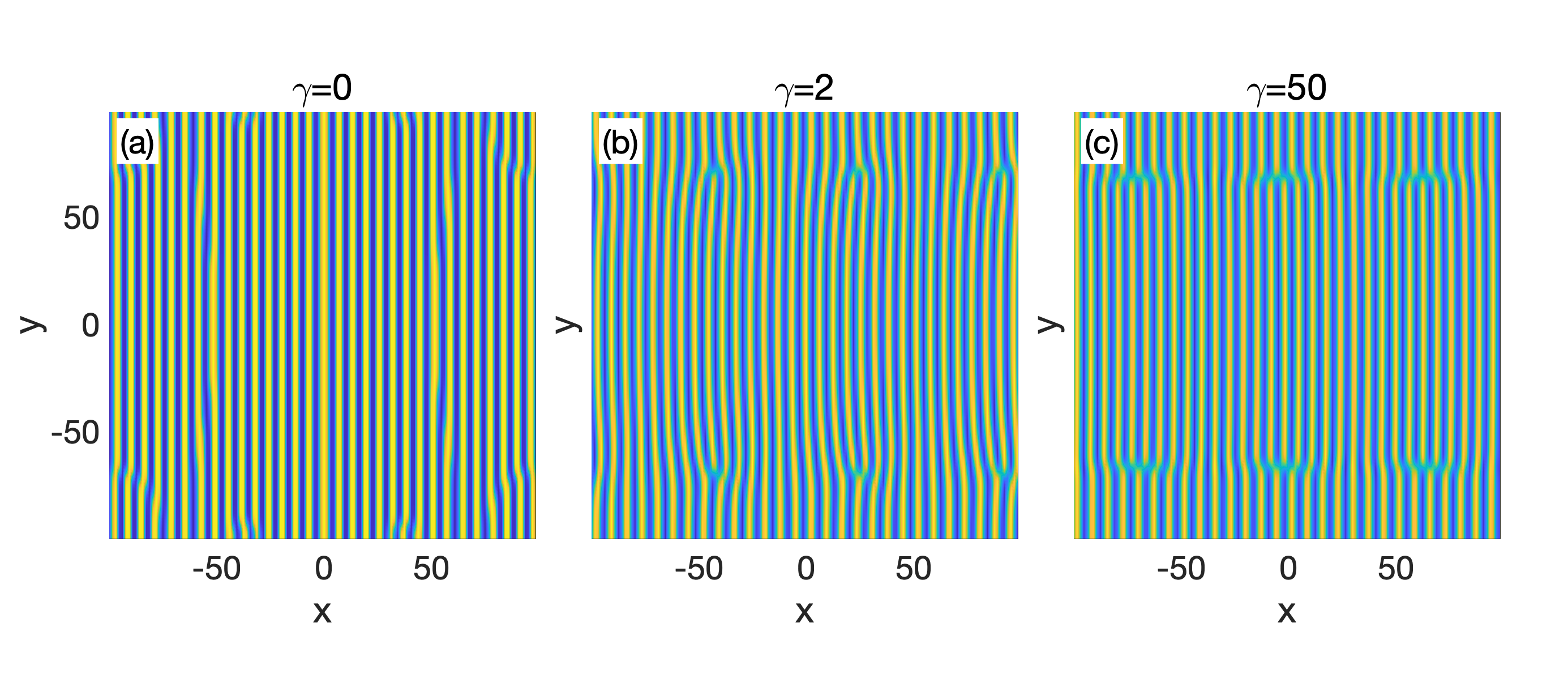}
    \caption{Solutions to Eq.~(\ref{eom2d}) on the domain \((x,y)\in [-100,100]^2\) are shown at time \(t=100\). The parameter values are \(\mu=\beta=1\) for each panel, and \(\gamma = 0\), 2, and 50, as labelled. The initial conditions were given by Eq.~ (\ref{eomchainsIC}) with \(k_1\) and \(k_2\) chosen so that \(n_1=28\) and \(n_2=31\). There are therefore \(n_2-n_1 = 3\) defects in each chain.}
    \label{fig:gamma_banded_ic}
\end{figure}

\begin{figure}
    \centering
    \includegraphics[width=\textwidth]{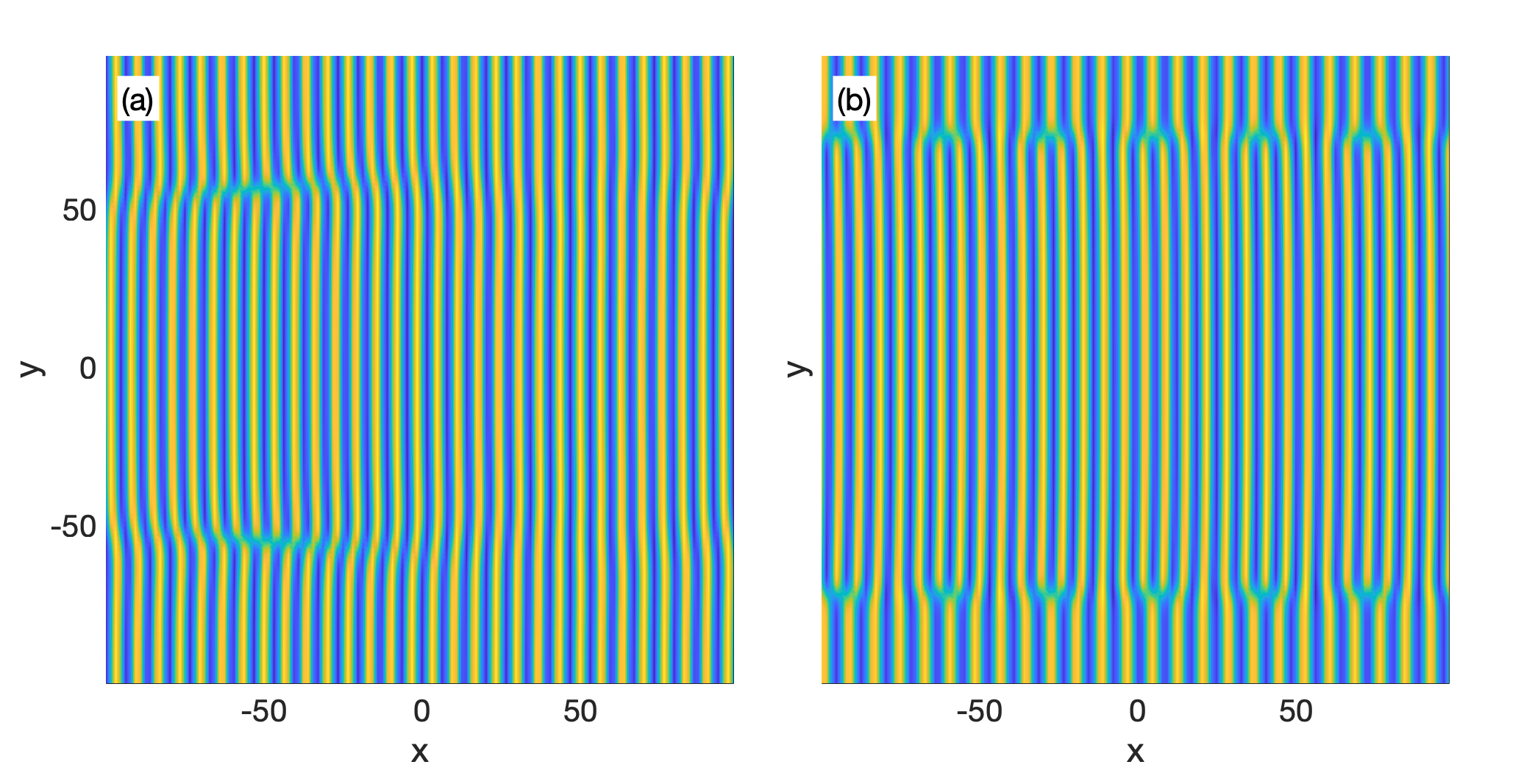}
    \caption{Solutions to Eq.~(\ref{eom2d}) on the domain \((x,y)\in [-100,100]^2\). The parameter values are \(\mu=\beta=1\) and \(\gamma=50\). The initial conditions were given by Eq.~ (\ref{eomchainsIC}).  Panel (a) shows a solution at time \(t=100\) with \(n_1=31\) and \(n_2=30\), and panel (b) shows a solution at time \(t=50\) with \(n_1=31\) and \(n_2=25\).}
    \label{fig:n_banded_ic}
\end{figure}

With a spatial white noise initial condition, chains of seams appear in an unpredictable fashion and the disordered arrangement of defects makes it challenging to discern the underlying order in the dynamics.  By choosing a different type of initial condition, we can produce defect chains in a controlled fashion that makes it easier to study them. In particular, we adopt an initial condition in which sinusoidal ripples of two different wave numbers \(k_1\) and \(k_2\) occupy horizontal bands and are in contact with one another: we set
\begin{align}
    \label{eomchainsIC}
    u(x,y,0)= 
    \begin{cases}
        \cos(k_1 x) & \text{ for } |y|<L/2 \,\text{ and } -L<x<L \\
        \cos(k_2 x) & \text{ for } |y|>L/2 \,\text{ and } -L<x<L.\\
    \end{cases}
\end{align}
The initial condition given by Eq.~(\ref{eomchainsIC}) must satisfy the periodic boundary conditions, and so we must have \(k_i = \pi n_i/L\), where \(n_i\) is an integer and \(i=1\) and 2.  We also choose \(k_1\) and \(k_2\) to be within the range of linearly unstable wave numbers, i.e., \((1-k_i^2)^2 <\mu\) for \(i=1\) and 2. This requirement ensures that neither of the initial sinusoids has an amplitude that rapidly tends to zero as time passes. Figure \ref{fig:eomchains}~(b) shows the result of a simulation with this type of banded initial condition.  Two defect chains have developed.  Notice that the dislocations within a defect chain all have the same sign and are evenly spaced. In addition, the dislocations in the two chains have opposite signs, and will annihilate after some time; see Fig.~\ref{fig:chain_annihilation}. Furthermore, as \(\gamma\) increases, the length of the defects increases, but is restricted by the number of defects in the chain (see Figs.~\ref{fig:gamma_banded_ic} and \ref{fig:n_banded_ic}).  Figure~\ref{fig:n_banded_ic}~(a) makes it particularly evident that that the seams are oriented obliquely to the \(x\) axis.

\subsection{Simulations of the Amplitude Equation}
\label{sec:ampsim}

\begin{figure}
    \centering
    \includegraphics[width=\textwidth]{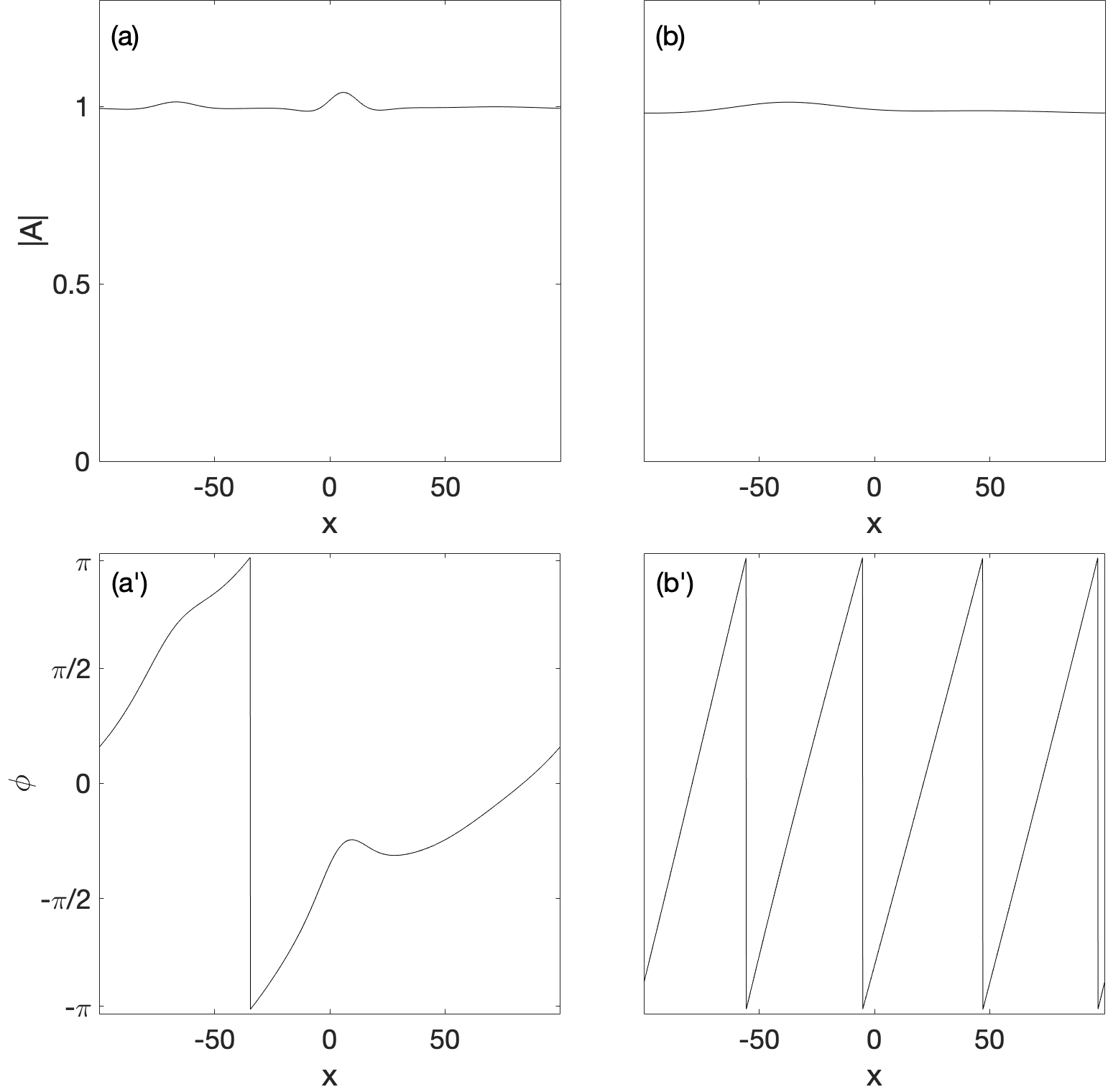}
    \caption{Two simulations of Eq.~(\ref{cgl1d}) starting from low amplitude spatial white noise initial conditions are shown at time \(t=60\). In panels (a) and (b), the amplitude \(|A|\) is plotted as a function of \(x\) for \(\eta = 10\) and \(100\), respectively. The corresponding phase \(\phi\) is depicted in panels (a') and (b').}
    \label{fig:cgl1d}
\end{figure}

Solutions of the 1D amplitude equation
\begin{align}
    \label{cgl1d}
    A_t = A + (1+i\eta)A_{xx} - |A|^2 A
\end{align}
behave in a fashion analogous to the solutions of the 1D DSHE (\ref{eom1d}). This is illustrated by the simulations of Eq.~(\ref{cgl1d}) shown in Fig.~\ref{fig:cgl1d}. The amplitude \(|A|\) and phase \(\phi\) are plotted as functions of \(x\) at time \(t=60\) for two simulations with \(\eta = 10\) and 100. For the larger value of \(\eta\), the solution is close to a plane wave: as seen in panels (b) and (b') of the figure, the amplitude \(|A|\) is almost a constant and the phase \(\phi\) is close to being a linear function of \(x\).  The plane-wave solution is the analog of the highly ordered ripples seen in Fig.~\ref{fig:1dsols} for the larger values of \(\gamma\).  The solution shown for \(\eta=10\) still deviates significantly from a plane wave at time \(t=60\) but approaches such a solution at longer times.

\begin{figure}
    \centering
    \includegraphics[width=\textwidth]{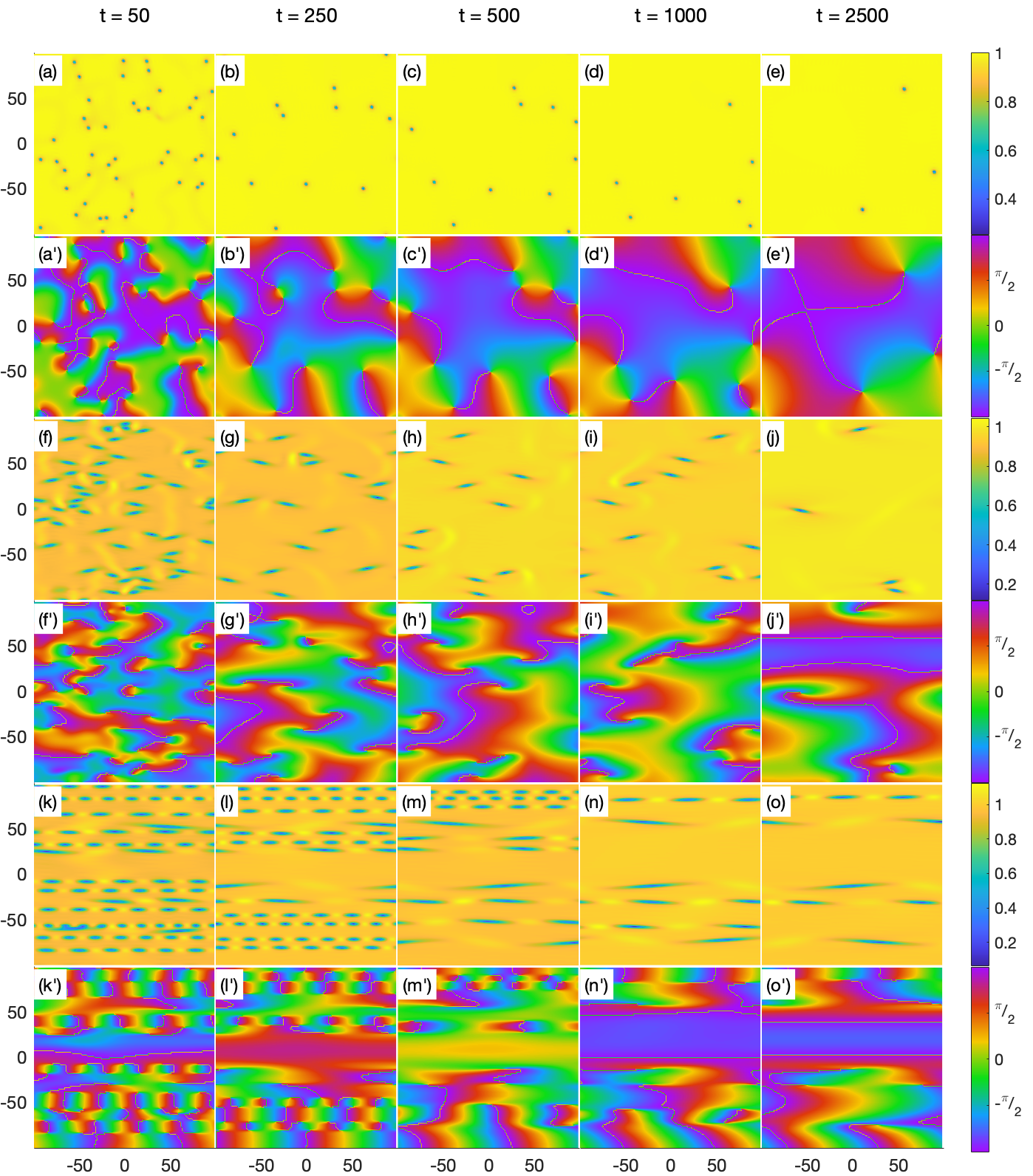}
    \caption{The time evolution of three simulations of Eq.~(\ref{standardamp}) on the spatial domain \((x,y)\in[-100,100]^2\). Rows (a)-(e), (f)-(j), and (k)-(o) show the magnitude of the solution \(|A(x,y,t)|\) for \(\eta=0,10\) and \(100\), respectively. Rows (a')-(e'), (f')-(j'), and (k')-(o') show the corresponding phases \(\phi(x,y,t)\). The solution at times \(t=50\), 250, 500, 1000 and 2500 is shown in columns 1 through 5, respectively.}
    \label{fig:ampsim}
\end{figure}

The analogy between the amplitude equation and the DSHE extends to 2D.
Figure \ref{fig:ampsim} shows simulations of Eq.~(\ref{standardamp}) at different times for selected values of \(\eta\). The initial condition in each case was low amplitude spatial white noise. For \(\eta=0\), Eq.~(\ref{standardamp}) reduces to the much studied real Ginzburg-Landau equation.

For \(\eta > 0\), the amplitude \(|A|\) is depressed in elongated regions that are obliquely oriented relative to the \(x\) direction, as seen in most clearly in Fig.~\ref{fig:ampsim} (f) - (j).  The phase \(\phi\equiv\Im(\ln A)\) winds through \(\pm2\pi\) about each of these regions.  These defects are the analogs of the seams in the DSHE and are spiral waves, as can be seen in panels (i') and (j') of Fig.~\ref{fig:ampsim}, for example.  The spiral waves are anisotropic, in contrast to the isotropic spiral waves produced by the (isotropic) complex Ginzburg-Landau equation.
As we would expect based on our simulations of the DSHE, chains of spiral waves appear in the simulations of the ACGLE (\ref{standardamp}).  
These are most evident in Fig.~\ref{fig:ampsim} (k) - (o). For \(\eta=0\), the spiral waves reduce to vortices.

\begin{figure}
    \centering
    \includegraphics[width=\textwidth]{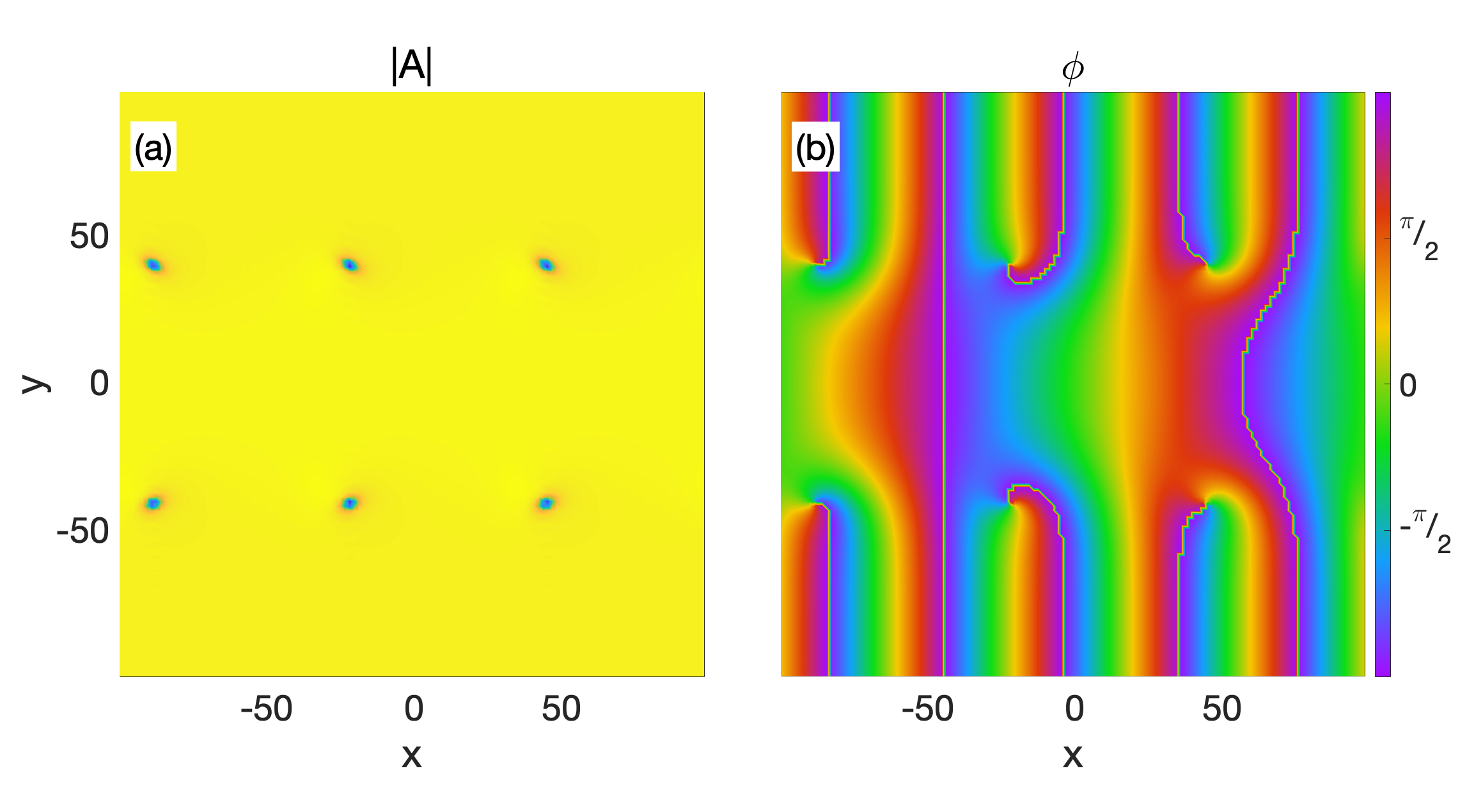}
    \caption{Chains of spiral waves created by simulating 
     Eq.~(\ref{standardamp}) with a banded initial condition of the form given by Eq.~(\ref{chainsIC}). The spatial domain was \((x,y)\in [-100,100]^2\) and the snapshot was taken at \(t=100\). We set \(\eta=1\) and the wave numbers \(q_1\) and \(q_2\) were chosen so that \(n_1=2\) and \(n_2=5 \).}
    \label{fig:spiral_chains}
\end{figure}

We can once again cause chains of defects to form in a controlled fashion using banded initial conditions.  We begin by noting that there is a plane-wave solution to Eq.~(\ref{standardamp}) of the form \(A(x,y,t)=R_0e^{i(q x-\omega t + \psi)}\), where \(R_0^2 = 1 - q^2\), \(\omega = \eta q^2\), and \(\psi\) is an arbitrary phase. We will study an initial condition that has two adjacent horizontal bands with different wave numbers \(q_1\) and \(q_2\) and phases \(\psi_1=\psi_2=0\):
\begin{align}
    \label{chainsIC}
    A(x,y,0)= 
    \begin{cases}
        \sqrt{1-q_1^2}e^{i q_1 x} & \text{ for } |y|<L/2 \,\text{ and } -L<x<L\\
        \sqrt{1-q_2^2}e^{i q_2 x} & \text{ for } |y|>L/2 \,\text{ and } -L<x<L.\\
    \end{cases}
\end{align}
The initial condition must satisfy the periodic boundary conditions, and so we must have \(q_i = \pi n_i/L\), where \(n_i\) is an integer and \(i=1\) and 2.
Without loss of generality,  we may assume that \(n_2>n_1\). Simulations with banded initial conditions show that two parallel chains of spiral waves form after a short time, as seen in Fig.~\ref{fig:spiral_chains} (a) and (b), for example.

If the plane waves simply propagated without changing their form, the solution to the ACGLE with the initial condition (\ref{chainsIC}) would be
\begin{align}
    \label{naive}
    A(x,y,t)= 
    \begin{cases}
        \sqrt{1-q_1^2}e^{i (q_1 x- \omega_1 t)} & \text{ for } |y|<L/2 \,\text{ and } -L<x<L\\
        \sqrt{1-q_2^2}e^{i (q_2 x- \omega_2 t)} & \text{ for } |y|>L/2 \,\text{ and } -L<x<L,\\
    \end{cases}
\end{align}
where \(\omega_i\equiv\eta q_i\) for \(i=1\) and 2.
This of course is not the solution to the initial value problem since the \(A(x,y,t)\) given by Eq.~(\ref{naive}) does not satisfy the ACGLE along the lines \(y=\pm L/2\).  Nevertheless, let us suppose for the moment that Eq.~(\ref{naive}) were the solution.  The defect cores would then appear at the locations \(x=x_n\) where the phase difference between the two bands is \(180^\circ\), i.e.,
\begin{align}
    q_1 x_n - \omega_1 t  = q_2 x_n - \omega_2 t  - (2n+1)\pi
\end{align}
for \(n \in \mathbb{Z}\). This would mean that
\begin{align}
    \label{vortloc}
    x_n = \frac{\omega_2-\omega_1}{q_2-q_1}t + \frac{(2n+1)\pi}{q_2-q_1}.
\end{align}
Equation (\ref{vortloc}) immediately gives us two results: 
the spiral wave velocity
\begin{align}
    \label{vortvel}
    \dot{x}_n = \frac{\omega_2-\omega_1}{q_2-q_1} = \eta(q_1+q_2)
\end{align}
and
the spacing between the cores of two adjacent spiral waves
\begin{align}
    \label{vortspacing}
    \Delta x = x_{n+1}-x_n = \frac{2\pi}{q_2-q_1}.
\end{align}
It is interesting to note that Eq.~(\ref{vortvel}) implies that \(\dot x_n\) is the sum of the phase velocities of the two plane waves.

As we have noted, Eq.~(\ref{naive}) does not really give the solution to the ACGLE with the banded initial condition. Instead, as time passes, the amplitude of the solution becomes depressed in the vicinity of the spiral wave cores and the lines of constant phase become curved, as Fig.~\ref{fig:spiral_chains} illustrates.  However, the initial condition (\ref{chainsIC}) is periodic in \(x\) with period \(\Delta x\).  As the solution to the ACGLE evolves in time, the solution remains periodic with this period.  Equation (\ref{vortspacing}) therefore gives the correct separation between the spiral wave cores.  In addition, our simulations demonstrate that Eq.~(\ref{vortvel}) gives a very good estimate of the spiral wave velocity, as we will now show.

\begin{figure}
    \centering
    \includegraphics[width=\textwidth]{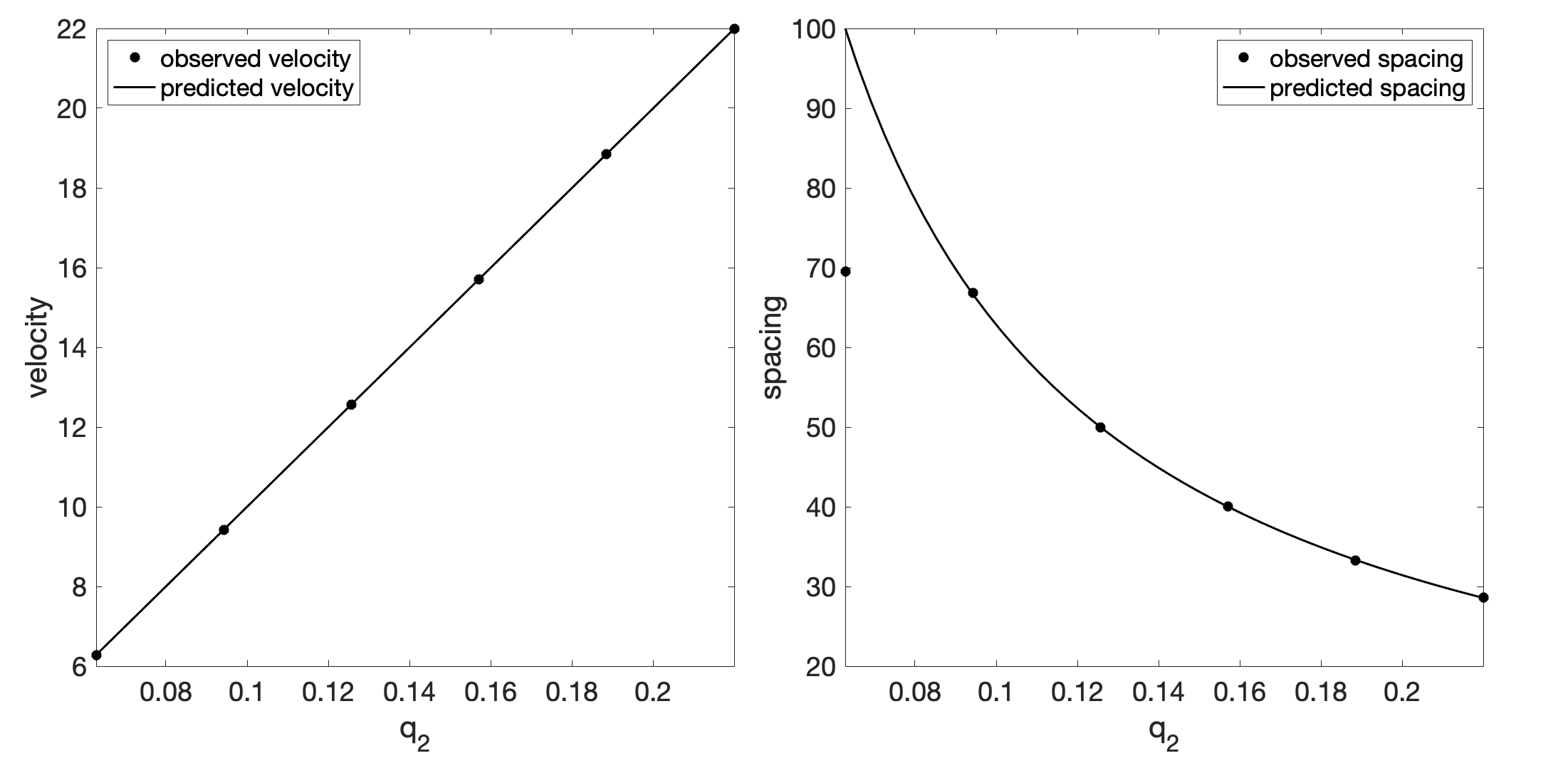}
    \caption{Simulations of Eq.~(\ref{standardamp}) with banded initial conditions of the form given by Eq.~(\ref{chainsIC}) were carried out 
    with \(q_1\) fixed at zero and with \(q_2=\pi n_2/L\), where \(n_2=2\), 3, 4, 5, and 7.  The spatial domain was \((x,y)\in [-100,100]^2\) and \(\eta\) was 100. The defect velocities and spacings were computed at \(t=200\). The observed values (dots) are compared to the values predicted by Eqs.~(\ref{vortvel}) and (\ref{vortspacing}) (solid curves).}
    \label{fig:defect dynamics}
\end{figure}

We compared the velocity and spacing predictions given by Eqs.~(\ref{vortvel}) and (\ref{vortspacing}) with the results of numerical simulations with banded initial conditions. Simulations were carried out for \(q_1=0\) and \(q_2=\pi n_2/L\), where \(n_2= 2, 3, 4, 5, 6\), and \(7\).  (We omitted the \(n_2=1\) case because the spacing between defects is undefined if there is only one defect in a chain.)  The simulations were performed for the parameter value \(\eta=100\) on the spatial domain \((x,y)\in[-100,100]^2\) and were run up to time \(t=200\).  The resulting defect velocities and spacings are compared with the predictions given by Eqs.~(\ref{vortvel}) and (\ref{vortspacing}) in Fig.~\ref{fig:defect dynamics} (a) and (b), respectively.  The agreement is excellent, provided that \(\eta\) and \(\Delta n\equiv n_2-n_1\) are sufficiently large.  If either \(\eta\) or \(\Delta n\) is too small, then the defects velocities oscillate in time.  This is the reason for the discrepancy seen in the right panel of Fig.~\ref{fig:defect dynamics} for the case \(n_2=2\).

\begin{figure}
    \centering
    \includegraphics[width=\textwidth]{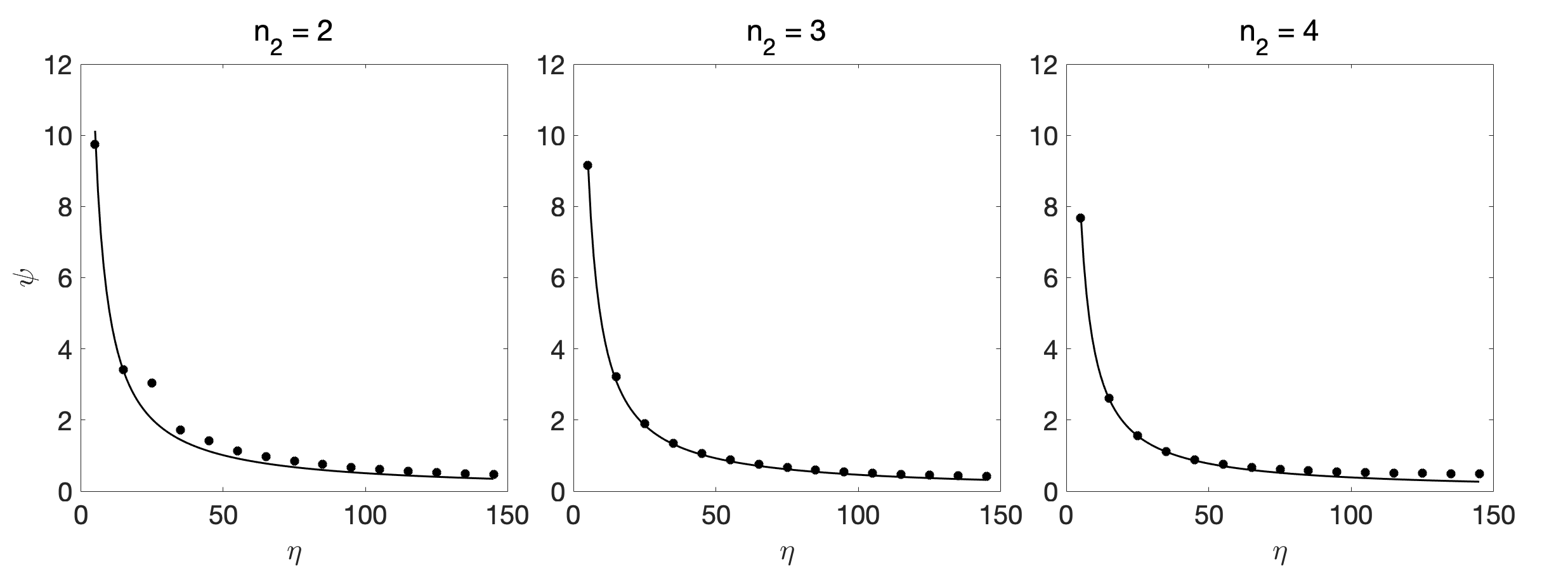}
    \caption{Simulations were run of Eq.~(\ref{standardamp}) with initial conditions given by Eq.~(\ref{chainsIC}) on the spatial domain \((x,y) \in [-100,100]^2\), and the average angle \(\psi\) that the defects made with the \(x\)-axis was computed. This was repeated for \(\eta = 5,10,\dots 145\) and for \(q_2 = n_2 \pi/100\) with \(n_2 = 2, 3,\) and \(4\). In each case, \(q_1=0\). Each data point represents the average angle obtained from a simulation, while the curve is a fit that is proportional to \(1/\eta\). The constant of proportionality depends on \(\Delta n\).}
    \label{fig:angle}
\end{figure}

A comparison of the regions of depressed amplitude \(|A|\) obtained for \(\eta=10\) and 100 in Fig.~\ref{fig:ampsim} suggests that the angle \(\psi\) that the spiral wave cores make with the \(x\) axis decreases with \(\eta\).  
To investigate this further, we defined a new function \(\rho\equiv 1-|A|^2\) within a neighborhood around a defect.  We interpreted \(\rho\) as a \lq\lq density,'' and then found the moment of inertia tensor for this density distribution. The angle that the principal axis with the smallest principal moment makes with the \(x\) axis is the angle \(\psi\). Figure~\ref{fig:angle} shows the value of \(\psi\) for a range of values of \(\eta\). The results are for banded initial conditions with \(q_1=0\) and \(q_2 = n_2 \pi/L\), where \(n_2=2\), 3 and 4. In addition, the values of \(\psi\) were averaged over all of the defects in a given simulation.  Our results support the proposition that \(\psi\) is a decreasing function of \(\eta\) for given values of \(n_1\) and \(n_2\).  They also suggest that \(\psi\) is inversely proportional to \(\eta\), and hence that \(\psi\) vanishes in the limit \(\eta\to\infty\).

Figure~\ref{fig:defect dynamics} shows that the defect spacing depends on \(q_2\), and of course it depends on \(q_1\) as well.  If we take the limit in which both \(q_1\) and \(q_2\) tend to a common nonzero value \(q\), then \(\Delta x\) tends to infinity according to Eq.~(\ref{vortspacing}).  In this limit, the seams are in effect infinitely wide and they become parallel to the \(x\) axis.  We found an exact solution of the ACGLE (\ref{standardamp}) that gives the form of the seams in this limit: 
\begin{align}
    \label{solA}
    A(x,y,t)=\pm\sqrt{1-q^2}e^{i(qx -\eta q^2 t)}\tanh\left(\sqrt{\frac{1-q^2}{2}}y\right).
\end{align}
Equation (\ref{solA}) is a valid solution for any real \(q\) with magnitude smaller than 1. If we cross the seam described by Eq.~(\ref{solA}) anywhere along its length, the phase \(\phi\) changes by \(\pi\).  The amplitude is depressed around the \(x\) axis in a region with width proportional to \((1-q^2)^{-1/2}\); this is the core of the seam.

\section{Discussion}
\label{sec:discussion}

This study was motivated in part by a need to better understand the nanoscale patterns produced by ion bombardment of solid surfaces.  Raised and depressed triangular regions that are traversed by ripples are commonly observed in experiments, but the formation of these patterns is not currently understood. 
Simulations of the dispersive KS equation in 2D produce triangular nanostructures that strongly resemble those seen in experiments and show that dispersion plays an important role in their genesis \cite{Loew2019}.  This finding led us to study the DSHE in 2D.

\begin{figure}
    \centering
    \includegraphics[width=\textwidth]{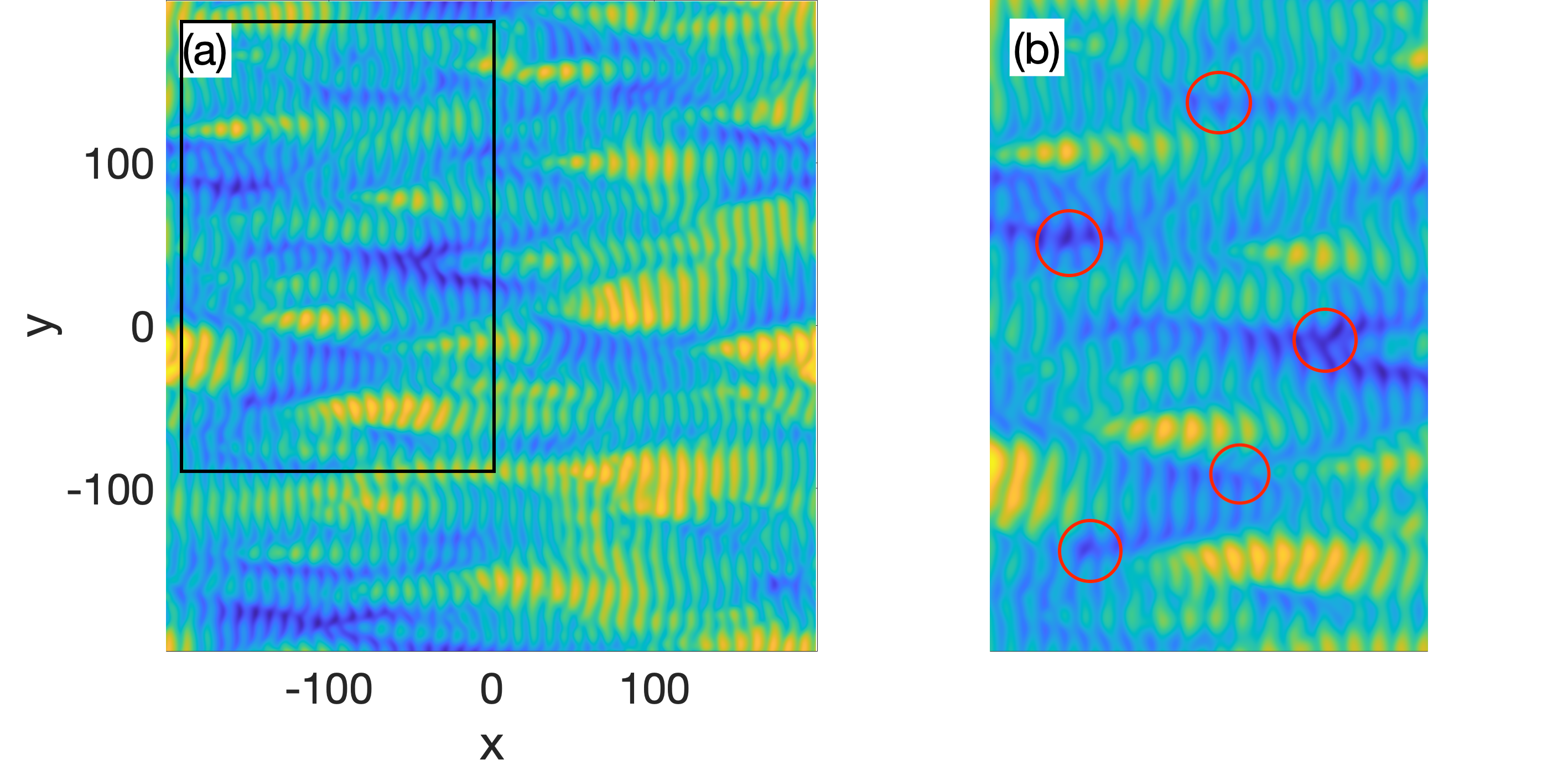}
    \caption{(a) A simulation of Eq.~(\ref{eom cr}) for \(\gamma=4\) at time \(t=85\) that shows the raised and depressed triangular regions traversed by ripples.  The initial condition was low amplitude spatial white noise. We employed an \(512 \times 512\) spatial grid and a time step of \(\Delta t = 0.1\). (b) An enlargement of the portion of panel (a) that is outlined in black.  The dislocation cores within five seams are circled.}
    \label{fig:my_label}
\end{figure}

Our work on the 2D DSHE suggests that the oblique sides of the triangular nanostructures might, in fact, be seams.  We therefore examined the results of a numerical integration of the simplified anisotropic KS equation with linear dispersion, Eq.~(\ref{eom cr}), and found that this is indeed the case.  This is illustrated by Fig.~\ref{fig:my_label}.  Our work therefore indicates that the notion that there are triangular nanostructures is misleading: Instead, the experimentally observed topographies are more properly thought of as ripples with a high density of seams.

The triangular structures found in simulations of the 2D dispersive KS equation are transient \cite{Loew2019}.  Because the surfaces display a high degree of disorder and the seams are abundant, it is challenging to discern how the so-called triangles disappear.  Our simulations of the dispersive KS equation and the associated amplitude equation suggest that seams of opposite signs move toward one another and then annihilate, ultimately leaving a surface without triangular nanostructures.

There are admittedly important differences between the dispersive KS equation and the DSHE in 2D.  The ripples are more orderly and the seams are more widely separated from one another in the case of the DSHE, for example. In addition, the anisotropic SHE we studied produces ripples with a high degree of order even in the absence of linear dispersion; in contrast, solutions of the anisotropic KS equation exhibit spatiotemporal chaos, and strong linear dispersion is needed to suppress this and to produce highly ordered ripples.  However, we exploited another key difference to our advantage.  The DSHE has small regions of unstable wave vectors near threshold which allowed us to derive the associated amplitude equation.  This is not possible in the case of the dispersive KS equation because there are unstable modes with arbitrarily long wavelengths.

It should be mentioned that the 1D DSHE (\ref{eom1d}) with the quadratic nonlinearity \(2u^2\) appended to the right-hand side has previously been studied \cite{Burke2009}.  The emphasis was on the propagation of fronts and on finding localized states for small $\gamma$, however.  In our work, we did not touch on those topics and considered only the case in which no quadratic nonlinearity appears in the equation of motion (\ref{eom1d}).  We also placed special emphasis on the limit in which the dispersive coefficient \(\gamma\) is large. 

Chains of spiral waves that appear in simulations of the ACGLE have been studied by Faller and Kramer \cite{faller99}.  Those authors had to carefully adjust the parameters in the ACGLE in order to get chains to form.  They also had difficulty getting chains of defects to form starting with spatial white noise initial conditions.  In this paper, we studied the special case of the ACGLE in which the coefficients of the terms proportional to \(A_{yy}\) and \(|A|^2 A\) are real.  In this case, chains of spiral waves form readily with a spatial white noise initial condition if linear dispersion is sufficiently strong.  We also established that chains of spiral waves can easily be produced in a controlled fashion using banded initial conditions.  This led us to a prediction of the spacing and velocity of the defects in a chain, and this prediction agrees well with our simulations.

\section{Conclusions}
\label{sec:conclusions}

Spatially extended dislocations were shown in this paper to appear in simulations of the 2D dispersive Swift-Hohenberg equation. These defects, which we call seams, tend to organize themselves into ordered chains.  The presence of a narrow band of unstable wavelengths in the DSHE allowed us to make analytical progress towards understanding seam defects.  We studied the DSHE in two limits.  First, close to threshold, we derived an amplitude equation for the DSHE, which turns out to be a special case of the ACGLE.  In this limit, seam defects correspond to spiral waves in the ACGLE.  Numerical simulations confirm analytical formulas for the distance between spiral wave cores and their velocities.  The second limit was that of large dispersion.  A perturbative analysis in this case yielded the propagation velocities of ripple patterns and a relationship between their amplitudes and wavelengths.  Our results shed light on the effect dispersion has on the nanoscale patterns produced by ion bombardment of solid surfaces.

\begin{acknowledgments}

This work was supported by Grants DMS-1814941 and DMR-2116753 awarded by the U.S.~National Science Foundation.

\end{acknowledgments}

\bibliography{bibtex_DSH}
\vfill\eject

\end{document}